\documentclass[twocolumn]{aastex62}
\pdfoutput=1 
\usepackage{amsmath,amstext}
\usepackage{CJK}
\usepackage[T1]{fontenc}
\usepackage[figure,figure*]{hypcap}
\usepackage{booktabs}
\usepackage{cleveref}
\usepackage{paralist}
\graphicspath{{fig/}}

\newcommand{\refsec}[1]{Section~\ref{#1}}
\newcommand{\reffig}[1]{Figure~\ref{#1}}

\newcommand{\refeqn}[1]{Equation~(\ref{#1})}

\newcommand{\fb}{f_\mathrm{b}}
\newcommand{\gq}{\gamma_q}
\newcommand{\BPRP}{ {G_{\rm BP}-G_{\rm RP} }}
\newcommand{\gaia}{\textit{Gaia} }
\newcommand{\G}{$G$ }
\newcommand{\Gbp}{$G_{\rm BP}$ }
\newcommand{\Grp}{$G_{\rm RP}$ }

\shorttitle{Modeling Binaries of OCs}
\shortauthors{Li \& Shao et al.}
\begin{document}
\begin{CJK*}{UTF8}{gbsn}

\title{Modelling unresolved binaries of open clusters in color-magnitude diagram. I. method and application of NGC
3532}
\correspondingauthor{Zhengyi Shao, Lu Li}
\email{zyshao@shao.ac.cn, lilu@shao.ac.cn}

\author[0000-0002-0880-3380]{Lu Li (李璐)}
\affil{Key Laboratory for Research in Galaxies and Cosmology, Shanghai Astronomical Observatory, Chinese Academy of
Sciences,
80 Nandan Road, Shanghai 200030, China.}
\affil{University of the Chinese Academy of Sciences, No.19A Yuquan Road, Beijing 100049, China.}

\author[0000-0001-8611-2465]{Zhengyi Shao (邵正义)}
\affil{Key Laboratory for Research in Galaxies and Cosmology, Shanghai Astronomical Observatory, Chinese Academy of
Sciences, 80 Nandan Road, Shanghai 200030, China.}
\affil{Key Lab for Astrophysics, Shanghai 200234, China}

\author[0000-0001-7890-4964]{Zhao-Zhou Li (李昭洲)}

\affil{Department of Astronomy, School of Physics and Astronomy, Shanghai Jiao Tong University,
955 Jianchuan Road, Shanghai 200240, China.}

\author{Jincheng Yu (俞锦程)}
\affil{School of Physics and Astronomy, Sun Yat-sen University, Zhuhai 519082, China.}

\author{Jing Zhong (钟靖)}
\affil{Key Laboratory for Research in Galaxies and Cosmology, Shanghai Astronomical Observatory, Chinese Academy of
Sciences,
80 Nandan Road, Shanghai 200030, China.}

\author{Li Chen (陈力)}
\affil{Key Laboratory for Research in Galaxies and Cosmology, Shanghai Astronomical Observatory, Chinese Academy of
Sciences,
80 Nandan Road, Shanghai 200030, China.}


\begin{abstract}

The binary properties of open clusters place crucial constraints on star formation theory and clusters' dynamical evolution. We develop a comprehensive approach that models the color-magnitude diagram (CMD) of the cluster members as the mixture of single stars and photometric unresolved binaries. This method enables us to infer the binary properties, including the binary fraction $f_\mathrm{b}$ and binary mass-ratio distribution index $\gamma_q$ when a power-law is assumed, with high accuracy and precision, which were unfeasible in conventional methods. We employ a modified Gaussian process to determine the main sequence ridge line and its scatter from the observed CMD as model input. As a first example, we apply the method to the open cluster NGC3532 with the \textit{Gaia} DR2 photometry. For the cluster members within a magnitude range corresponding to FGK dwarfs, we obtain $f_\mathrm{b} = 0.267\pm0.019$ and $\gamma_q = - 0.10\pm0.22$ for binaries with mass ratio $q > 0.2$. The $f_\mathrm{b}$ value is consistent with the previous work on NGC3532 and smaller than that of field stars. The close to zero $\gamma_q$ indicates that the mass ratios of binaries follow a nearly uniform distribution. For the first time, we unveil that the stars with smaller mass or in the inner region tend to have lower $f_\mathrm{b}$ and more positive value of $\gamma_q$ due to the lack of low mass-ratio binaries. The clear dependences of binary properties on mass and radius are most likely caused by the internal dynamics. 

\end{abstract}


\section{Introduction}
\label{sec:intro}

Open clusters (OCs) are excellent laboratories for the study of binary systems. First, it is generally accepted that almost all stars formed in clusters \citep{Lada2003}, and most of them are thought to form in binary or multiplicity systems (\citealt{Duquennoy1991}, \citealt{Goodwin2005}, \citealt{Kouwenhoven2007a}, \citealt{Rastegaev2010}). Although most of the clusters will eventually be dissolved into field stars, some remain as OCs as we can observe now. Therefore, the properties of binaries in OCs provide essential constraints on the star formation scenario. Second, binaries deeply involve in the dynamical evolution of stellar clusters. In the early stage, according to the Heggie-Hill law \citep{Heggie1975}, close encounters in stellar clusters involving binary systems may disrupt {\it soft} (i.e., generally wide) binaries efficiently (\citealt{Heggie1975}, \citealt{Kaczmarek2011}, \citealt{Reipurth2012}, \citealt{Li2013}, \citealt{deGrijs2015}, \citealt{Deacon2020}). Only {\it hard} (close) binaries remain. Subsequently, in the central (dense) region of a cluster, with the frequent encounters, {\it hard} binaries may get harder due to the kinematic energy exchanges. This process actually provides a gravitational fuel that can delay and eventually stop the gravitational collapse, though it mostly happens in globular clusters \citep{Binney2011}.

Moreover, the OCs cover wide ranges of age and chemical abundance. Their properties, e.g., the binary fraction $\fb$ and the mass-ratio distribution, together with their evolutions, can be easily compared for different OCs with ages and metallicities, which are strictly related to the environments of the Milky Way. Therefore, binaries are ideal tracers for investigating the formation and evolution of OCs, as well as the dependence on their forming locations and environments. Furthermore, binary stars are essential in determining the total mass \citep{Borodina2019a} and the stellar mass function \citep{Kroupa2002a} of OCs.

Most binaries are unresolvable in images. Only some can be directly identified by measuring variations of their radial velocity~\citep{Mateo1996} and/or luminosity~\citep{Milone2012}. Both approaches are limited to bright targets with relatively high orbital inclinations and short periods, leading to significant velocity changes or at least partial eclipses caused luminosity varies.

In contrast to these low discovery efficiency and {\it expensive} binaries, detecting the unresolved binaries in the color-magnitude diagram (CMD) are much {\it cheaper}. They mostly lay on the brighter and redder direction of the main sequence (MS) of single stars in the CMD. In particular, equal mass binaries appear exactly at 0.75 mag brighter than their MS counterparts. Thus, the binary properties can be inferred from the statistical analysis based on CMD. In this paper, we focus on those photometric unresolved binaries, and call them binaries for short in the following context.

The most general way to study binaries in the CMD is simply dividing the CMD into single and binary regions and counting cluster members in each region respectively (see e.g., \citealt{Sollima2010}, \citealt{Clem2011b}, \citealt{Milone2012}, \citealt{Li2013} and references therein for applications in star clusters). However, it is non-trivial to infer the mass ratio distribution. Meanwhile, it is only a rough estimation of the binary fraction because the observational errors will blend the single stars and the low mass-ratio binaries.

Instead of classifying each cluster member as a single star or a binary, a better treatment is to model the CMD as a mixture of both components, which is characterize by the $\fb$ and the mass-ratio distribution. Such idea was first adopted by \citet{Hurley1998} to generate synthetic cluster and by \citet{Naylor2006}, \citet{Kalirai2004} to fit the isochrone with the presence of binaries. There are numerous investigations on binary fraction of OCs by comparing the observations with the synthetic CMDs (\citealt{Bonifazi1990}, \citealt{Sarro2014}, \citealt{Sheikhi2016b},\citealt{Li2017}), however, they have not made attempt to measure the mass ratio distribution simultaneously. The mass ratio distribution of binaries is also a fundamental quantity. It may reflect the physics in binary formation, e.g., pairing mechanics \citep{Kouwenhoven2009b}, also shed light on the subsequent dynamical evolution. Moreover, the mass ratio distribution can affect interpretation of observation, e.g., the cluster mass determination that depends on various mass ratio distributions, including the "luminosity-limited pairing" as called by \cite{Borodina2019a}.

In this work, we propose a comprehensive approach based on the mixture model to establish a continuous number density distribution in the CMD and infer the properties for binaries. It not only can measure the binary fraction and mass ratio distribution simultaneously, but also can conveniently take into account the photometric error and/or the intrinsic dispersion by convolving the model with appropriate scatters.

A practical issue about this method is that the inference of these binary parameters is sensitive to the exact location and extension of the MS. However, it is reported that the theoretical isochrones show a small but significant deviation from the observed MS \citep{Fritzewski2019}. Such an amount of deviation might significantly affect the inferred binary properties. It is necessary to take a small correction to the theoretical isochrone according to the empirical MS. We employ a modified Gaussian process to determine the MS ridge line empirically and estimate its scatter from the observed CMD as model input (see more discussions in \refsec{sec:method:MM:NumDen}).

On the other hand, the field star contamination is a long-standing problem of OCs. Field stars in the CMD may severely affect the determination of binary properties. Thanks to the release of {\gaia} data \citep{GaiaCollaboration2018b}, especially the sufficiently precise astrometric data including the high-quality proper motion, which allows us to separate the cluster members from the field stars effectively \citep[G18]{GaiaCollaboration2018f}. Thus, for those OCs with better membership determination, we only need to consider the distribution of members. Besides, the {\gaia} DR2 provides accurate photometric data of passbands \G, \Gbp and \Grp \citep{Evans2018}. So the \gaia data is optimal for the OC studies. In this work, we test the expected performance of our method with mock samples using the \gaia passbands and corresponding precision, and apply to NGC3532 with real data.

Please note that, binarity is not the only effect that causes the stars to deviate from MS. The existence of triple or higher-order systems (\citealt{Mermilliod1992}, \citealt{Kouwenhoven2007a}) may also behave a deviation like binaries. They are usually wide systems and seem to have a higher frequency in younger stellar clusters or associations. These young higher-order systems are unstable and will be significantly disrupted by internal dynamics (\citealt{Reipurth2012}, \citealt{Elliott2016}). Thus, the fraction of higher-order multiplicities decreases rapidly with age, which leads to a negligible effect in OCs. Another effect is the optical blending, which may degenerate with the binary detection in globular clusters. However, it is not important for OCs due to the relative low spatial density.

The rest of the paper is organized as follows. In Section~\ref{sec:method}, we demonstrate how to construct a mixture model in the CMD to constrain the properties of unresolved binaries. Meanwhile, a modified Gaussian process is developed to determine the empirical MS of a cluster. In Section~\ref{sec:Mock}, we validate our method with a set of mock samples. In Section~\ref{sec:3532}, the method is applied to open cluster NGC3532 with a {\it main-sample} that includes FGK dwarfs and subsamples based on stellar mass or central radius. The comparison of the binary parameters and their implication on the dynamical process are discussed in Section~\ref{sec:discu}, followed by the conclusion summarized in Section~\ref{sec:sum}.

\section{Model and algorithm}
\label{sec:method}

The strategy of our approach is using the mixture model method, which is a natural statistical method for many situations in astronomy\citep{Kuhn2017a}. It is verified that even in a heavily mixed case, like the parallax distributions of field stars and members of distant globular clusters \citep{Shao2019}, the mixture model method can also fit the distribution parameters precisely.

In this work, the mixture model is applied for the star distribution in the CMD. The observed cluster members, including single stars and binaries, are isolated points, whereas the models will be established as smoothed number density distribution models. To build up each model distribution, we first need to determine the MS of single stars, which can either be a theoretical isochrone or an empirical ridge line taken from a modified Gaussian process (Section~\ref{sec:method:EMS}). We also determine the binary sequences (BSs) with different mass-ratios. Then, based on these sequences, together with the stellar mass function, the binary fraction, and the mass-ratio distribution, we compute the model number density by combining both of single star and binary components (Section~\ref{sec:method:MM:NumDen}). We further evaluate the scatters of magnitude and color, and convolve them to mimic the real observation. Finally, the joint likelihood of the sample stars that are belonging to a specific model is calculated, and used to infer the concerned parameters.

\subsection{Mixture model of cluster members}
\label{sec:method:MM}

\subsubsection{Locations of star on CMD}
\label{sec:method:MM:iso}

The members of an open cluster formed in the same molecular cloud simultaneously as a single stellar population (SSP). They are practically at the same distance with very similar chemical composition and dust extinction. Single stars are located exactly on an isochrone line in the CMD and vary with stellar mass ($\mathcal{M}$). The majority of them will stay below the turn-off points for a long time, hence present as a MS apparently. Therefore, the MS is the fiducial location of an observed cluster in the CMD.

An unresolved binary is two stars that are too close to be resolved in the image. That means, they will appear as a single point source with the light of their summation. Suppose their masses are $\mathcal{M}$ for the primary star and $q\mathcal{M}$ for the secondary one, where $q$ is the so called mass-ratio of this binary within $0 < q \leqslant 1$. We further assume that the binary is a detached system, and these two stars have been evolving independently. Thus, in any photometric passband, the magnitude of this binary is the simple combination of these two single stars,
\begin{equation}\label{eq:mb}
m_{\rm b}(\mathcal{M},q) = -2.5\log[10^{-0.4m_{\rm s}(\mathcal{M})}+10^{-0.4m_{\rm s}(q\mathcal{M})}],
\end{equation}
\noindent where $m_{\rm s}(\mathcal{M})$ and $m_{\rm s}(q\mathcal{M})$ are magnitudes of the primary and secondary stars. The color index, $c_{\rm b}(\mathcal{M},q)$, is then calculated from corresponding passbands. In particular, $m_{\rm b} \simeq m_{\rm s} - 0.75$ when $q=1$.

\begin{figure}[!htbp]
  \centering
  \includegraphics[width=0.45\textwidth]{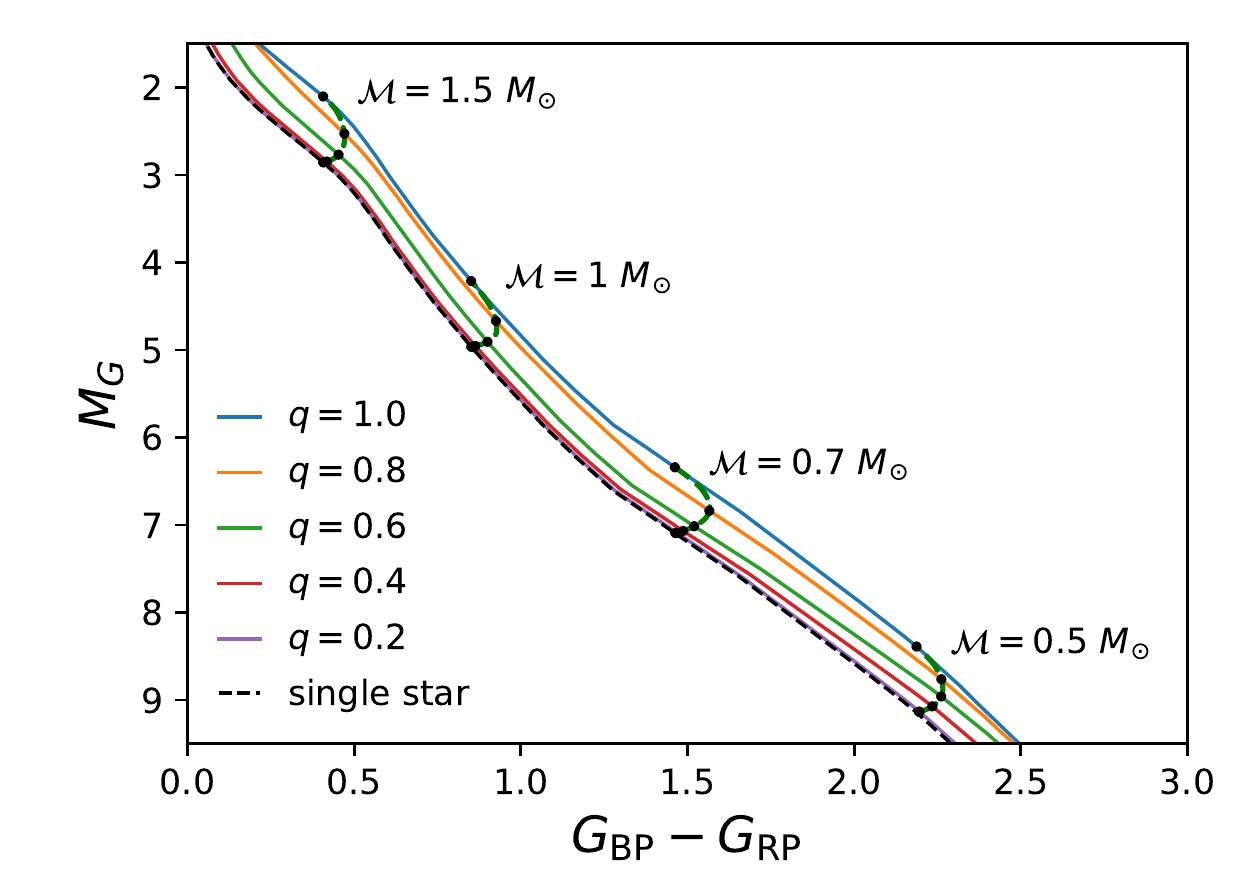}
  \caption{Main sequence (MS) and binary sequences (BSs) of the {\gaia} DR2 photometric data, \G and $\BPRP$. The stellar population is set to be 400\,Myr with solar metallicity. Solid lines show the BSs with $q=0.2, 0.4, 0.6,0.8$ and 1.0 respectively. The dashed black line is the MS of single stars for comparison. Dashed short green lines mark the binaries with primary mass ($\mathcal{M}$) equal to $0.5,0.7,1.0,1.5\,M_\odot$ for various $q$ values. The BS of $q=0.2$ is very close to the MS with the discrepancy in color to be $\sim 0.0092$ or $\sim 0.0012$ mag for $\mathcal{M}=0.5$ or $1.5 \,M_\odot $ respectively.}
  \label{fig:binSeq}
\end{figure}

In \reffig{fig:binSeq}, we generate a curve of cluster member binaries having a constant $\mathcal{M}$ but with different mass-ratio (dashed short green lines). In cases of $q < 1$, the binary appears to be brighter and redder than its single primary star. On the other hand, we can generate a line for a fixed $q$ value with different primary mass $\mathcal{M}$. This constructs one of the BSs. Especially, the equal-mass binaries ($q=1$) form a 0.75 mag brighter BS which is in parallel to the MS. It is clear that all these cluster members, both of the single stars and the detached binaries, locate in a belt range constrained by the MS and the equal-mass BS. The width is about 0.15 mag in the color ($c=\BPRP$) at $ \mathcal{M} \sim 1\,M_\odot$.

It should be mentioned that the BSs are non-equal separated in the CMD. For a small value of mass-ratio, e.g., $q=0.2$, the BS is very close to the MS, since the secondary star is too faint to affect the total luminosity and color. The differences in magnitude and color are $\Delta G = -0.0078$ mag, $\Delta c = 0.0092$ mag for $\mathcal{M}=0.5 M_\odot$ and $\Delta G = -0.0006$ mag, $\Delta c = 0.0012$ mag for $\mathcal{M}=1.5 M_\odot$. These values are smaller than the typical observational error of \gaia and may be much smaller than the extension of the MS of a real cluster (see Section~\ref{sec:method:MM:NumDen} and ~\ref{sec:method:EMS:ext} for details and Table~\ref{tab:MSRL} of NGC3532 as an example). So it is difficult to distinguish the low mass-ratio binaries from the single stars. Fortunately, in the mixture model method, $\fb$ is a {\it global} fitting parameter that avoids the problem of distinguishing. Although the contribution of low mass-ratio binaries keeps uncertain, it will be restricted by the mass-ratio distribution function, which is mainly curved by the higher mass-ratio binaries.\\

\begin{figure*}[!htbp]
  \centering
  \includegraphics[width=1\textwidth]{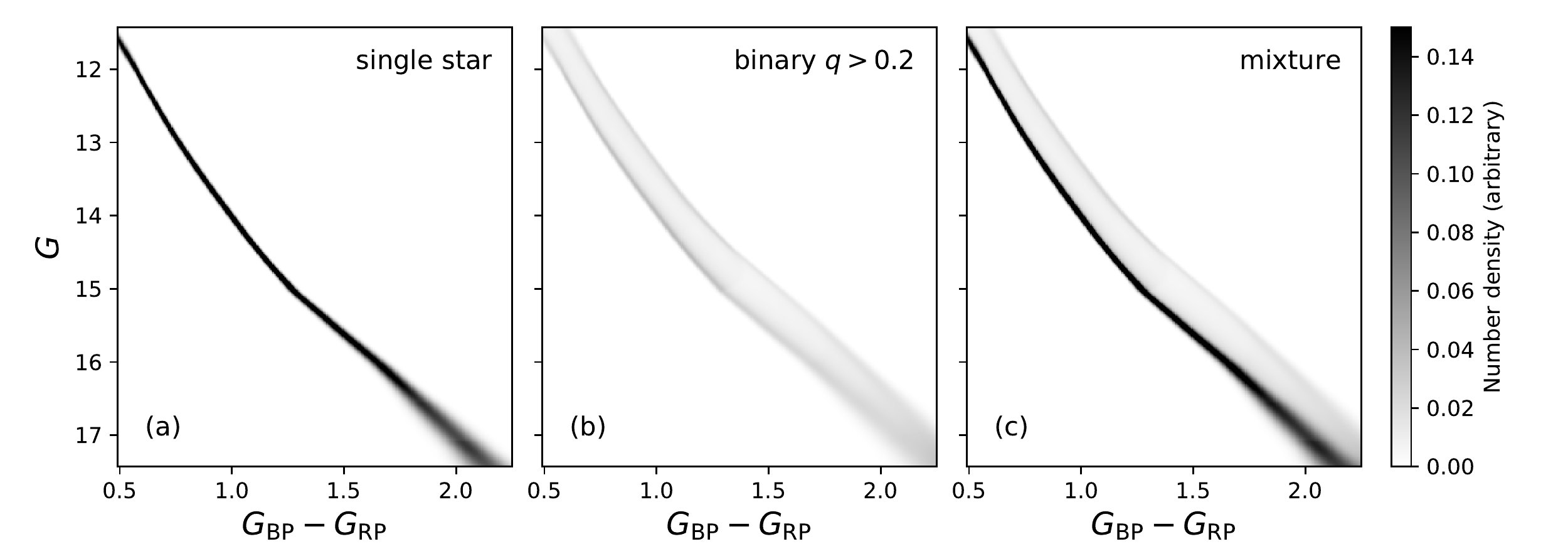}
  \caption{The model number density distribution of single stars (a), unresolved binaries (b) and their mixture (c).
 For illustration, we use a SSP at age = 400Myr with solar metallicity and adopt a mass function with power index $-2.35$. For binaries, we adopt $f_{\rm b}=0.27$, $\gq=0.0$ and $q_{\rm min}= 0.2$. The scatters, $\sigma_m$ is 0.01 magnitude and $\sigma_c$ is a function of $G$ which listed in Table \ref{tab:MSRL} .
 }
  \label{fig:mimo}
\end{figure*}

\subsubsection{Number densities in the CMD}
\label{sec:method:MM:NumDen}

\noindent\textbf{Single star population}

We first define a model number density $\rho_{\rm s}(m,c)$ to describe the MS of single stars in the CMD. We use the symbol $\Theta_\mathrm{SSP}$ to refer any constraints characterizing the MS, it can be either the SSP parameters for a theoretical isochrone, such as age, metallicity, distance, and extinction,
or an non-parametric empirical MS as discussed in \refsec{sec:method:EMS} and \ref{sec:3532:gener:MSRL}. Actually, $\rho_{\rm s}$ is a $\delta$-function at $(m_{\rm s},c_{\rm s})$, where the magnitude $m_{\rm s}(\mathcal{M}|\Theta_\mathrm{SSP})$ and the color $c_{\rm s}(\mathcal{M}|\Theta_\mathrm{SSP})$ are functions of $\mathcal{M}$ for a given SSP ($\Theta_\mathrm{SSP}$). We omit $\Theta_\mathrm{SSP}$ in following equations for simplicity. Then, the number density contributed by a star on MS writes
\begin{equation}\label{eq:rho-s}
\rho_{\rm s}(m, c |\mathcal{M}) = \delta [m - m_{\rm s}(\mathcal{M}), c - c_{\rm s}(\mathcal{M})].
\end{equation}
\noindent Given the stellar mass function $\mathcal{F}_\mathrm{MF}$, the number density of the MS for the cluster is described as
\begin{equation}\label{eq:phi-s}
\phi_\mathrm{s}(m,c)= \int \rho_{\rm s}(m, c |\mathcal{M} )\mathcal{F}_\mathrm{MF}(\mathcal{M})d\mathcal{M}. \\
\end{equation}

\noindent\textbf{Unresolved binaries}

Assuming a binary system contains two stars with masses $\mathcal{M}$ and $q\mathcal{M}$. The magnitude and color are $m_{\rm b}(\mathcal{M},q)$ and $c_{\rm b}(\mathcal{M},q)$. Then, the BS with a fixed value of $q$ can also be expressed as a $\delta$-function,
\begin{equation}\label{eq:rho-b}
\rho_{\rm b} (m, c |\mathcal{M},q) = \delta [ m - m_{\rm b}(\mathcal{M},q), c - c_{\rm b}(\mathcal{M},q)].
\end{equation}
Considering the mass function $\mathcal{F}_\mathrm{MF}$, as well as the mass-ratio distribution $\mathcal{F}_q(q)$, the number density of binaries is
\begin{equation}\label{eq:phi-b}
\phi_\mathrm{b}(m,c)= \int_{q} \int_{\mathcal{M}} \rho_{\rm b} (m, c |\mathcal{M},q )\mathcal{F}_\mathrm{MF}(\mathcal{M})
\mathcal{F}_q(q) d\mathcal{M} dq. \\
\end{equation}

\noindent\textbf{The forms of $\mathcal{F}_\mathrm{MF}$ and $\mathcal{F}_q$ }

In this work we use the single power-law mass function of Salpeter \citep{Salpeter1955}
\begin{equation}\label{eq:F-MF}
\mathcal{F}_\mathrm{MF} (\mathcal{M}) = \frac{dN}{d\mathcal{M}} \propto \mathcal{M}^{-\alpha_{\rm MF}},
\end{equation}
which is found to be suitable for NGC3532 (See discussion in Section~\ref{sec:3532:MF}). Since this cluster is middle-aged, and particularly considering that only the mass range of [0.5,1.5]\,$M_\odot$ will be used, as clearly shown in \reffig{fig:3532-MF}, we do not need to adopt the break power-law shape. Nevertheless, it is possible to use other forms of mass function depending on the specific cluster in concern.

The binary mass-ratio is usually assumed to follow a power-law distribution as well (\citealt{Kouwenhoven2007a}, \citealt{Reggiani2013}, \citealt{Duchene2013}),
\begin{equation}\label{eq:F-q}
\mathcal{F}_q (q)= \frac{dN}{dq} \propto q^{\gamma_q}. \\
\end{equation}
\noindent Thus, the power index $\gamma_q$ is the only parameter that shapes the mass-ratio distribution. A positive $\gamma_q$ means that there are more massive secondary stars, and vise versa.
As we will discuss in \refsec{sec:discu:gamma},
other forms of mass ratio distribution have been considered in literature (\citealt{Fisher2005,Kouwenhoven2007a,Kouwenhoven2009b,El-Badry2019,Borodina2019a}, see \citet{Duchene2013} for a comprehensive summary).
It is straightforward to use any alternative forms in our model.\\

\noindent\textbf{The mixture model and fitting parameters}

The number density for the whole cluster is the summation of contributions of single stars and binaries,
\begin{equation}\label{eq:phi-mix}
  \phi(m,c) = (1-f_{\rm b}) \phi_\mathrm{s}(m,c) + f_{\rm b} \phi_{\rm b}(m,c).
\end{equation}
It is fully described by the stellar population properties of $\Theta_\mathrm{SSP}$ and $\mathcal{F}_{\rm MF}$, the binary features of
$\fb$ and $\mathcal{F}_q$, and could be written as $ \phi(m,c \,|\, \Theta_\mathrm{SSP}, \alpha_{\rm
MF},\fb,\gq)$.

For the purpose of this paper, we only focus on the binary properties: the binary fraction and the mass-ratios. Therefore we assume that other parameters or features are all well-determined. Thus, the only parameters that we need to fit are $f_{\rm b}$ and $\gamma_q$. \\

So far, $\phi(m,c \,|\,\fb,\gq)$ is the intrinsic model distribution for the whole cluster without any extension caused by observational errors or other dispersion factors.
\\

\noindent\textbf{Extension of the sequences }

To mimic the real observational data in the CMD, the theoretical distribution $\phi$ should be further broadened by the
observational errors of magnitude and color. Besides, other factors may also extend the sequences. For example, the extra extension may be caused by cluster's age spread (\citealt{Palla2005} found about 10 Myrs in Orion Nebula Clusters). Other effects like the small variation of the metallicity of cluster members, or the unknown inhomogeneous extinction from dust in the cluster, the change of luminosity or color due to the star rotation, or even the underestimation of the photometric errors. Each factors may act a complicated spread of MS, with some of them are asymmetry. In practice, also based on the central limit theorem, we use one single Gaussian scatter with the standard deviation of $\sigma_c$ in color to approximate the accumulation of all these possible broadens (see Section~\ref{sec:method:EMS:ext} for how to determine the magnitude dependent scatters empirically from observation data). For the BSs, we assume that they have the same broadening in the CMD as that of the MS. Therefore, the apparent model distribution $\psi$ is a convolution of the theoretical $\phi$ and the Gaussian kernel of $(\sigma_m,\sigma_c)$, where $\sigma_m$ and $\sigma_c$ are the characteristic scatters at given magnitudes. We have
\begin{align}\label{eq:psi}
    \psi(m,c\,|\,\sigma_m,\sigma_c,\fb,\gq) = & \int \phi(m',c'\,|\,\fb,\gq) \nonumber \\
                                    & \cdot \mathcal{N}(m,c\,|\,m',c',\sigma_m,\sigma_c )\,dm'dc' .
\end{align}
\noindent where $\mathcal{N}(m,c\,|\,m',c',\sigma_m,\sigma_c )$ represents the Gaussian probability centered on $(m',c')$. Now the apparent model distribution $\psi$ is ready to be compared with observational data points on CMD for fitting the parameters.

The details of the adopted empirical MS and its extension will be discussed in Section~\ref{sec:method:EMS} and~\ref{sec:3532:gener:MSRL}. \\

\noindent\textbf{Digitization of $\psi$}

To build the practical models of $\psi(m,c\,|\,\sigma_m,\sigma_c,\fb,\gq)$, we compute the digitized {\it map} of the number density in the CMD with various values of $\gamma_q$ and $f_{\rm b}$. According to the difference between the MS and the low mass-ratio BS ($q=0.2$, $\Delta c \lesssim 0.01$ mag), and also considering the dispersion of the real cluster NGC3532 with $\sigma_m \lesssim 0.01$ mag and $\sigma_c \sim 0.007$ to $0.07$ mag (see table~\ref{tab:MSRL}), we use the grid sizes of 0.005 mag for both $\Delta m $ and $\Delta c$ on CMD.

\reffig{fig:mimo} shows an example of the model distribution of single stars, binaries and their mixture. In this case, even though the binary mass-ratio is set to be uniformly distributed ($\gamma_q=0.0$), it is clear that the number density looks like congregating in either higher or lower mass ratio locations. This is another expression of the non-equidistant sequences as shown in Figure~\ref{fig:binSeq}. In terms of the observation, it can partly explain the phenomena that many clusters show obvious BS $\sim 0.75$ mag above the MS. \\

\noindent\textbf{Magnitude limitations}

Practically, we have to constrain the magnitude range of the sample. The faint end is usually truncated by the observational flux limit. Meanwhile, a cut-off at the bright end might also be necessary. Considering that the stellar properties are possibly more complicated near the turn-off point, e.g., the MS broadening due to stellar rotation \citep{Maeder1974}, we should adopt a magnitude fainter than the turn-off point as the bright limit. For a sample restricted in the magnitude range $[m_1,m_2]$, we can further re-normalize $\psi$ by a factor
\begin{equation}\label{eq:C}
    \mathcal{C} = \left[ \int_{c}\int_{m_1}^{m_2} \psi(m,c) \,dm dc \right]^{-1}.
\end{equation}
It is worth mentioning that the binary fraction derived from our method is the {\it global} value of the underlying population. In contrast, direct counting the number of binaries within a magnitude range suffers to a sample selection effects and likely overestimates the binary fraction. Since the binaries have smaller primary masses compared to the single stars of the same magnitude, the counted binary fraction will be higher than that of the whole population
when the smaller stars are more abundant due to the negative mass function power index. This counting bias motivates people (e.g., \citealt{Milone2012}) to cut the data along the equal $\mathcal{M}$ line on the CMD (see Figure~\ref{fig:binSeq}). Anyway, the above issue is irrelevant to our method. The sample selection will not affect any results once it has been modeled into the re-normalization factor. It will be further justified by mock tests in Section~\ref{sec:Mock:rst}. We prefer the simple cuts on magnitude for convenience since the photometric survey's data quality is usually a function of magnitude. Additionally, it could include more sample stars than the equal $\mathcal{M}$ cut.

\subsubsection{Likelihood and parameter inference}
\label{sec:method:MM:LH}

We suppose that the likelihood of the $i$th star follows the apparent model number density of distribution in the CMD, which is the chance
that it appears at the given point of ($m_i,c_i$),
\begin{equation}\label{eq:lh}
    \mathcal{L}_i \propto \mathcal{C}\psi_i = \mathcal{C}(\gamma_q,f_{\rm b})\, \psi(m_i,c_i\,|\,\fb,\gq),
\end{equation}
\noindent then write down the joint likelihood for the whole sample as
\begin{equation}\label{eq:lhtot}
    \mathcal{L}(\mathbf{m},\mathbf{c}|\,\fb,\gq) = \prod \mathcal{L}_i(m_i,c_i|\,\fb,\gq),
\end{equation}
\noindent where $(\mathbf{m},\mathbf{c})$ denote the set of magnitude and color for all sample stars.

According to the Bayesian inference framework, the {\it posterior} probability density function (PDF) is
\begin{equation}\label{eq:pdf}
    P(\fb,\gq|\,\mathbf{m},\mathbf{c}) \propto \mathcal{L}(\mathbf{m},\mathbf{c}|\,\fb,\gq)\cdot \pi(\fb,\gq),
\end{equation}
\noindent where $\pi(\fb,\gq)$ is the {\it prior} PDF of the parameters, which are set to be flat within $[0,1]$ for $\fb$ and $(-\infty,+\infty)$ for $\gq$ respectively.

We employ the Monte Carlo Markov Chain (MCMC) with the public package \texttt{emcee} \citep{Foreman-Mackey2013} in sampling of the fitting parameters to obtain their {\it posterior} PDF. Then, we compute the marginal PDF for each of them, and in this paper, throughout use the 50\% position, half of the 16\% to 84\% width to represent the fitting results and their uncertainties.

Besides these two fitting parameters, we can also compute the marginal PDF for other derived parameters, like $\fb^{0.5}$ or $\fb^{0.7}$, by using the MCMC sampling records of each pair of $(\fb, \gq)$ with the corresponding $\mathcal{F}_q(q|\gq)$. Then, their results and uncertainties are calculated in the same way as those of original fitting parameters.

Practically, we have to set a minimum value of $q$, and use $\fb^{q_{\rm min}}$ instead of $\fb$ as the fitting parameter. That means, all the $q<q_{\rm min}$ binaries will be treated as single stars in the model. It is used to avoid the possible divergence of $\mathcal{F}_q(q|\gq)$ at $q=0$ when $\gq < 0$. Moreover, the theoretical isochrones, like the PARSEC, usually have the lower limit of the stellar mass ($\sim 0.08 M_\odot$), which will also make a constrain of the lowest mass of the secondary star that we can evolved. In this work, considering the minimum $\mathcal{M}$ will be used is $0.5 M_\odot$, we choose $q_{\rm min}=0.2$, and still use the symbol $\fb$ to represent the $\fb^{0.2}$ for short in the rest part of the paper. In this case, other fractions, like $\fb^{0.5}$ for $q>0.5$, $\fb^{0.7}$ for $q>0.7$, and $\fb^\mathrm{tot}$ for all binaries.\\

\subsection{Empirical main sequence and extension}
\label{sec:method:EMS}

Although the stellar evolution models have been remarkably improved in recent years, the theoretical isochrones do not always match the observed MS of a real cluster to a satisfactory level. The discrepancy could be due to the imperfectness of the stellar evolution theory or of the observational photometric calibration. Unfortunately, the resulting $\fb$ and $\gq$ are very sensitive to the color of stars, since the low mass-ratio binaries (e.g. $q < 0.2$) are very close to the MS. For example, if the theoretical isochrone is only 0.02 mag bluer than the real MS, all single stars will be misregarded as small $q$ binaries, while $\gq$ will be biased to a more negative value. Noticing that, observationally, there is usually a distinct ridge line (RL) of the number density indicating the location of the MS, a natural solution is to use the RL as an empirical MS (hereafter MSRL) instead of using the theoretical one.

It is not a trivial task to find the MSRL of the observational data. The easiest way is to draw it on the CMD by hand \citep{Fritzewski2019}. Alternatively, one can calculate the peak of the histogram of the color distribution for each magnitude bins, just like \citet{Milone2012} fitted the MSRL of a globular cluster. Unfortunately, both of these two methods are not optimal for OCs due to the limited number of member stars and hence the large Poisson noise.

In fact, this task can be regarded as a robust regression problem with outliers. Therefore, we propose a new robust regression method based on the Gaussian process (see e.g., \citealt{Rasmussen2006}) with the iterative trimming to locate the ridge of color as a function of magnitude, and subsequently derive the scatter in color ($\sigma_c$). The details will be given in a separate paper (Li et al.\ 2020 in preparation)\footnote{The Python code is publicly available at \url{https://github.com/syrte/robustgp}}. We briefly introduce the procedure in the following subsection.
\\

\subsubsection{Robust Gaussian process}
\label{sec:method:EMS:GP}

The Gaussian process regression is a powerful machine-learning algorithm that allows us to fit a smooth function from noisy data without binning or assumptions in the parametric form. However, the standard Gaussian process might give biased outputs when the sample contains non-Gaussian outliers, e.g., the additional BSs in our case. We try to remove the outmost outliers in an iterative way to minimize their influence in determining the ridge. We first run the Gaussian process with the full sample to derive the expected ridge of color as a function of magnitude. Then we remove a subset of stars that have the largest separation to this mean function and rerun the Gaussian process with the remaining sample to update the ridge function.\footnote{Note that these stars are always selected from the full sample, i.e., a star discarded in an earlier iteration might be taken back later.} We repeat the procedure until the convergence. Although the predicted ridge function of the first several iterations might deviate significantly from the real MS due to the contamination of the binaries, it converges to the final ridge fast and robustly. The precision of this method will be testified with mock clusters in Section~\ref{sec:Mock:MSRL}.

We also note that a similar treatment based on iterative trimming to outliers in color was adopted by \cite{Clem2011b}. We might expect that our MSRL is more precise and continuous thanks to the advantages of the Gaussian process.

\subsubsection{Measurements of the MSRL extension}
\label{sec:method:EMS:ext}

As we find that the extension of the MS is usually larger than the typical uncertainties of the observational color at a given magnitude, we have to measure the extension independently. For each magnitude bin, we calculate the standard deviation ($\sigma_c$) of the residual color of the MSRL. To avoid the contamination of binary stars, we only use the stars on the blue side and double them by reflecting over the MSRL. After that, we smooth these discrete $\sigma_c$ values as a continued function of magnitude. This method will also be testified with the mock data (Section~\ref{sec:Mock:MSRL}). \\

\section{Validation with mock cluster}
\label{sec:Mock}

Here we aim to test the validity and accuracy of our mixture model for estimating the binary properties. We will constrain the mock clusters to be similar to the real cluster NGC3532. We would expect almost same performance for clusters with different ages and metallicities. Because inferring the binary properties only uses the information that how binaries distribute away from the MS, not the shape of the MS.

We use the fiducial parameters of NGC3532 that listed in Table~\ref{tab:3532-basic} for mock clusters, including the age, metallicity, distance, dust extinction, and mass function parameter. We set parameters of the binaries to be $f_{\rm b}= 0.27$ and $\gamma_q=0.0$, which is also similar to the fiducial results of NGC3532. The scatter of magnitude $\sigma_m$ is assumed to be 0.01 mag, which are the typical value of \G band of the {\gaia} DR2 and the scatters of color $\sigma_c$ are also set as a function of magnitude as what we have derived from NGC3532 (see Table~\ref{tab:MSRL}). Moreover, $f_{\rm b}$ and $\gamma_q$ will be varied to test for different cases, and the $\alpha_{\rm MF}$ will be changed to discuss the influence of mass function.\\

\subsection{Mock procedure}
\label{sec:Mock:proc}

\begin{figure*}[!htbp]
  \centering
  \includegraphics[width=1.0\textwidth]{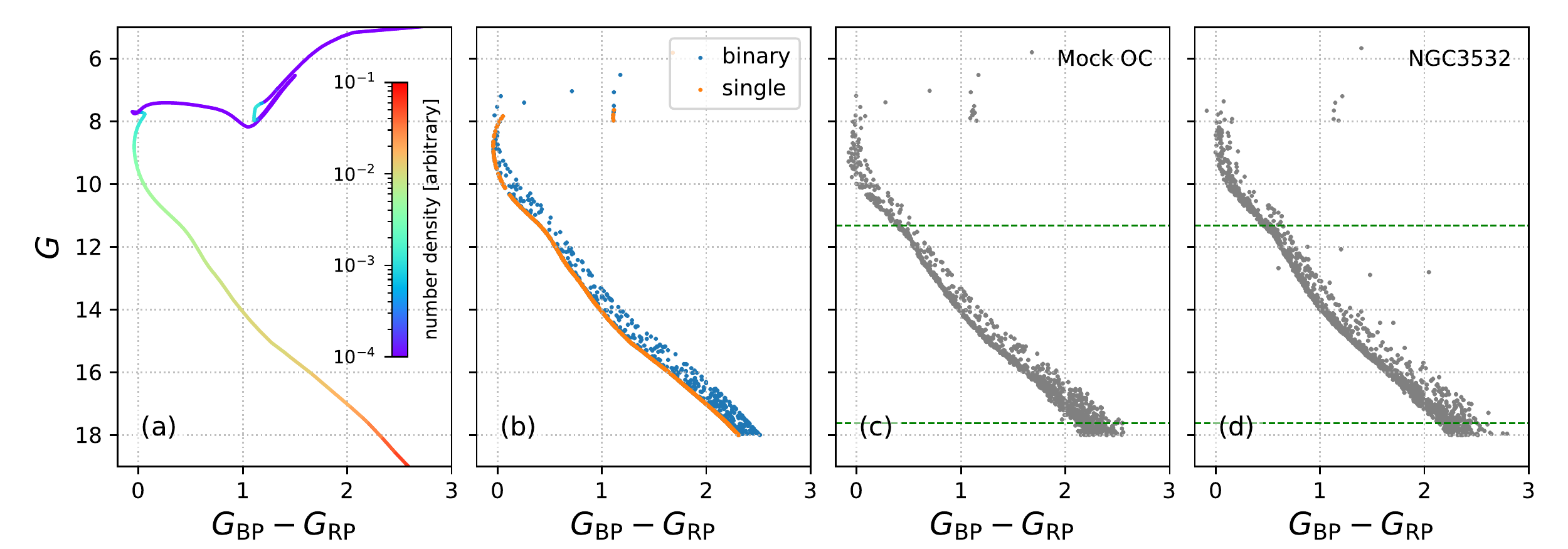}
  \caption{Schematic for generating a mock cluster. (a) The theoretical number density of a SSP from PARSEC isochrone for given parameters same as NGC3532. The mass function index $\alpha_{\rm MF}=-2.35$; (b) The distribution of mock cluster members in CMD, with $\fb=0.27$ and $\gq=0.0$; (c) Each mock star is finally perturbed by $\sigma_m$ and $\sigma_c$, while the truncation is the magnitude limitation of the data; (d) The CMD of real NGC3532 members for comparison. In panels (c) and (d), the stars enclosed by two dashed lines represent the {\it main sample} with $\mathcal{M}=0.5$ to 1.5 $M_\odot$ for single stars.
  }
  \label{fig:gener-mock}
\end{figure*}

The mock procedure is similar to \citet{Perren2015} and schematized in Figure~\ref{fig:gener-mock}. In the mock cluster, we randomly assign a mass $\mathcal{M}$ to each star following the mass function distribution (\refeqn{eq:F-MF}). Then, a fraction $f_{\rm b}$ of them are assumed as binaries with primary star mass unchanged while secondary star mass of $q \mathcal{M}$, where $q$ is randomly chosen from the mass-ratio distribution $\mathcal{F}_q(q|\gamma_q)$. Subsequently, we derive their magnitudes and colors according to the theoretical isochrone (PARSEC \citep{Bressan2012} \footnote{PARSEC version 1.2S, \url{http://stev.oapd.inaf.it/cgi-bin/cmd}} characterized by the cluster age, metallicity with the {\gaia} photometric system ($G$, $G_\mathrm{BP}$ and $G_\mathrm{RP}$, \citealt{Evans2018}). Binaries are then calculated by the combination of two corresponding stars (Equation~\ref{eq:mb}).

To mimic the real observation, we convert the absolute magnitude and the intrinsic color to the apparent magnitude and the reddened color with the given distance and dust extinction. We further give them random small shifts according to $\sigma_{m}$ and $\sigma_c$. Moreover, we define the {\it main sample} as the same as what will be adopted for NGC3532 having the flux limitation of $[11.32, 17.62]$ mag, corresponding to the single star mass between $[0.5, 1.5]\,M_\odot$ (see Section~\ref{sec:3532:MainSamp}). After this procedure, the total number of member stars is about $N \sim 1400$, which is comparable to the real case of NGC3532.

Figure~\ref{fig:gener-mock} panel (c) shows the final distribution of the mock cluster, which is very similar to the real observation data of NGC3532 in panel (d). \\
 \begin{figure}[!htbp]
  \centering
  \includegraphics[width=0.5\textwidth]{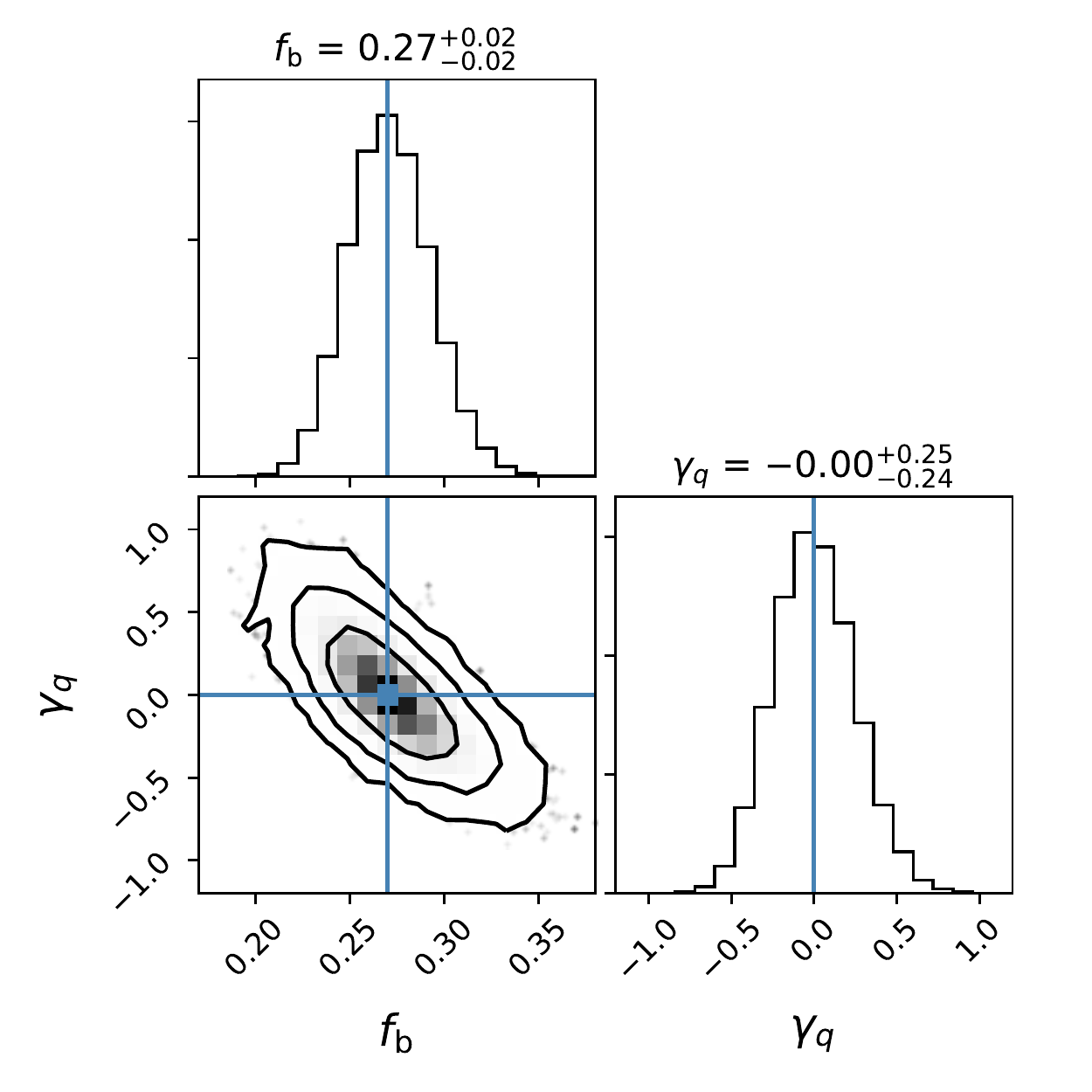}
  \caption{
  Probability density function (PDFs) of parameters $\fb$ and $\gq$ based on the \texttt{emcee} sampling of an example mock cluster. The contours correspond to the $1\sigma$, $2\sigma$ and $3\sigma$ (68.3\%, 95.4\% and 99.7\%) confidence levels, whereas the blue lines indicate the true values. Histograms show the marginalized probability distributions for estimating of $\fb$ and $\gq$, respectively.
  }
\label{fig:single-mcmc}
\end{figure}

\subsection{General performance of fitting results}
\label{sec:Mock:rst}

The fitting results of an example mock cluster are shown in Figure~\ref{fig:single-mcmc}. Clearly, the inferred values of $f_\mathrm{b}$ and $\gamma_q$ are in good agreement with the true values, and their uncertainties are remarkably small. There is a slight degeneracy between $\fb$ and $\gq$, which appears as the smaller $\gq$ leading to the larger $\fb$. This is comprehensible since the uncertainty is mainly due to the low mass-ratio binaries, and the overestimation of them will cause a more negative value of $\gq$.
{\color{blue}
\begin{table}[htbp]
\caption{Fitting results of $\fb$ and $\gq$ for 100 mock clusters.
}
\begin{center}
\setlength{\tabcolsep}{10pt}
\begin{tabular}{rccrc}
\midrule \midrule
  Cases &  $f_{\rm b}$ & $\sigma_{\fb}$ & $\gamma_{q}$ & $\sigma_{\gq}$\\
\midrule
\multicolumn{5}{l}{\it  fiducial case } \\
   (a) & 0.268 & 0.018  &  0.03 & 0.21 \\
\midrule
\multicolumn{5}{l}{\it  different luminosity ranges } \\
   (b) &  0.264& 0.029  &  0.06 & 0.33 \\
   (c) &  0.274& 0.031  &  -0.01 & 0.37 \\
\midrule
\multicolumn{5}{l}{\it  shifted isochrones } \\
   (d) &   0.998 & -  &  -1.81 & 0.06 \\
   (e) &  0.161& 0.009  &  1.25 & 0.16 \\
\midrule
\multicolumn{5}{l}{\it  variants of $\sigma_c$ } \\
   (f) &  0.409& 0.016  &  -0.82 &  0.11 \\
   (g) &  0.233   &   0.019    & 0.36   & 0.25    \\
\midrule
\multicolumn{5}{l}{\it  MSRL } \\
   (h) &   0.265& 0.028  &  0.07 & 0.23 \\
\midrule \midrule
\end{tabular}
\end{center}
\tablecomments{--- The $\fb$ and $\gq$ are shown as their median values of 100 mock clusters. The true values are $\fb=0.27$ and $\gq=0.0$. The dispersions of $\sigma_{\fb}$ and $\sigma_{\gq}$ are calculated as the half-width of the 16\% to 84\% distribution ranges. Different cases: (a) general results for the {\it main sample}; (b) the brighter half of the {\it main sample} with $ G < 15.06 $ mag; (c) the fainter half of the {\it main sample} with $G > 15.06$ mag; (d) the fiducial MS is shifted towards blue with $\Delta c = -0.02$ mag, and (e) shifted towards red with $\Delta c = 0.02$ mag; (f) using the typical observational errors of $\BPRP$ from the {\gaia} DR2 instead of the given model extension $\sigma_c$ of the MS; (g) artificially enlarge the $\sigma_c$ to 1.2 times of the given values; (h) same as the case (a) but using the MSRL instead of the PARSEC isochrone. }
\label{tab:mock-rst}
\end{table}
}

By using 100 mock clusters, we systematically investigate the fitting results for different conditions. The median values of the best fitting results and their dispersions are summarized in Table~\ref{tab:mock-rst}. For the fiducial case of the {\it main samples}, both of $f_{\rm b}$ and $\gamma_{q}$ are very close to the true input values of the mock, while their dispersions are at the same level of their fitting uncertainties. Such consistency is also shown in the cases of brighter or fainter subsamples except for a bit larger of dispersion which is due to the decrease of the sample size.

When we directly count $N_{\rm b}(q\geqslant0.2)$ for the {\it main sample} of mock clusters, we find $\langle N_{\rm b}/N \rangle = 0.287$, which is obviously larger than the given binary fraction 0.27 due to the negative power index of the mass function, as we have mentioned in Section~\ref{sec:method:MM:NumDen}. However, this effect is mostly eliminated in the fitting results of $\fb$. It justifies our approach can detect the {\it global} value of $\fb$ that is independent with the magnitude range we adopted.

Other cases are assumed if there are biases in measuring of the fiducial MS of a mock cluster. For instance, if it is 0.02 mag bluer than the true place (case (d)), almost all stars are regarded as binaries, and the $\gq$ becomes more negative since there is a large increasing number of the low mass-ratio binaries. On the contrary, if the MS is 0.02 mag redder (case (e)), $\fb$ will significantly decrease and $\gq$ will increase, due to many low mass-ratio binaries are misregarded as single stars. Moreover, if the MS is not shifted, but its extension is underestimated, e.g., we use the {\gaia} observational errors on color to represent the whole scatter (case (f)), one may find that there are also more single stars to be disregarded as binaries, while the $\gq$ tends to have more negative value correspondingly. Furthermore, if we used the overestimated $\sigma_c$, e.g., artificially enlarge them to 1.2 times of the measured MS's extension, we can find that more small $q$ binaries are misregarded as single stars. Although the change of $\fb$ is small in this case, the $\gq$ increases due to the missing of small $q$ binaries. These results reiterate the importance of the accuracy of the MS and its extension for a real cluster.\\

\begin{figure}[!htbp]
  \centering
  \includegraphics[width=0.5\textwidth]{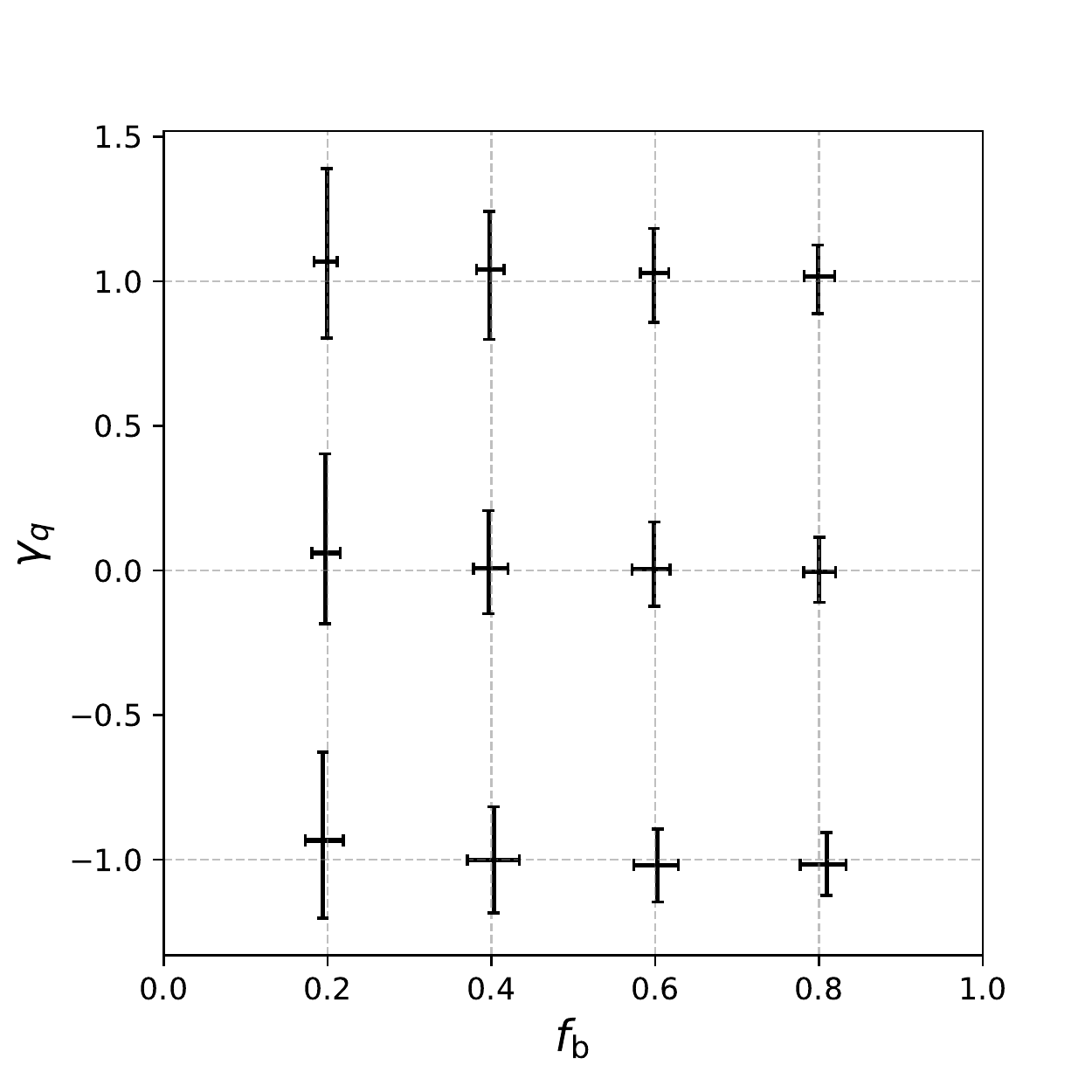}
  \caption{
The precision of fitting results of $\fb$ and $\gq$ for different sets of mock samples. We take the true values of $f_\mathrm{b} = 0.2, 0.4, 0.6, 0.8$ and $\gamma_q = -1.0, 0.0, 1.0$, and the results are shown as their median values and corresponding dispersions of 100 mock clusters for each set. }
  \label{fig:mock-rst}
\end{figure}

\subsection{Varying of $f_{\rm b}$ and $\gamma_q$ }
\label{sec:Mock:Vary}

We apply our method to 12 sets of mock clusters with different $\fb$ and $\gq$ to cover a wide range of these two parameters. We take $f_\mathrm{b} = 0.2, 0.4, 0.6, 0.8$ and $\gamma_q = -1.0, 0.0, 1.0$ respectively, but keep the other cluster parameters unchanged with at their fiducial values. For each choice of ($\fb$, $\gq$), we make 100 mock clusters.

The results are summarized in \reffig{fig:mock-rst}. The error bars show the corresponding dispersions of every group of 100 mock clusters. Overall, our method provides unbiased estimation for both parameters. There are two systematic variations. First, $\sigma_{\gq}$ is increased with a decreasing of $\fb$. One may also find that the $\sigma_{\fb}$ is increased with the decreasing of $\gq$, though they are all very slight. These two trends can both be attributed to the difficulty of separating low mass-ratio binaries by single stars. In all, we can conclude that our approach provides an effective estimation of binary properties with both accuracy and acceptable precision.\\

\subsection{Influence of mass function}
\label{sec:Mock:MF}

In this work, we adopt the \citet{Salpeter1955} mass function with the power-law index of $-2.35$. However, the slope of the mass function is not universal for different clusters and even for different ranges of mass or radius for the same clusters, depending on the initial condition and evolution stage. The complexities are not able to be captured by a simple power-law mass function. Motivated by the potential influence of imperfect mass function, here we apply our method to 5 sets of mock clusters with $\alpha_\mathrm{MF}={-3, -2.5, -2.35, -2.0, -1.5}$ respectively, while keeping the other cluster parameters at their fiducial values. We make 100 mock clusters for each choice of $\alpha_\mathrm{MF}$, but conformably adopt $\alpha_\mathrm{MF}\equiv-2.35$ in the distribution model for fitting to mimic the case of misuse of the mass function profile.

\begin{figure}[!htbp]
  \centering
  \includegraphics[width=0.5\textwidth]{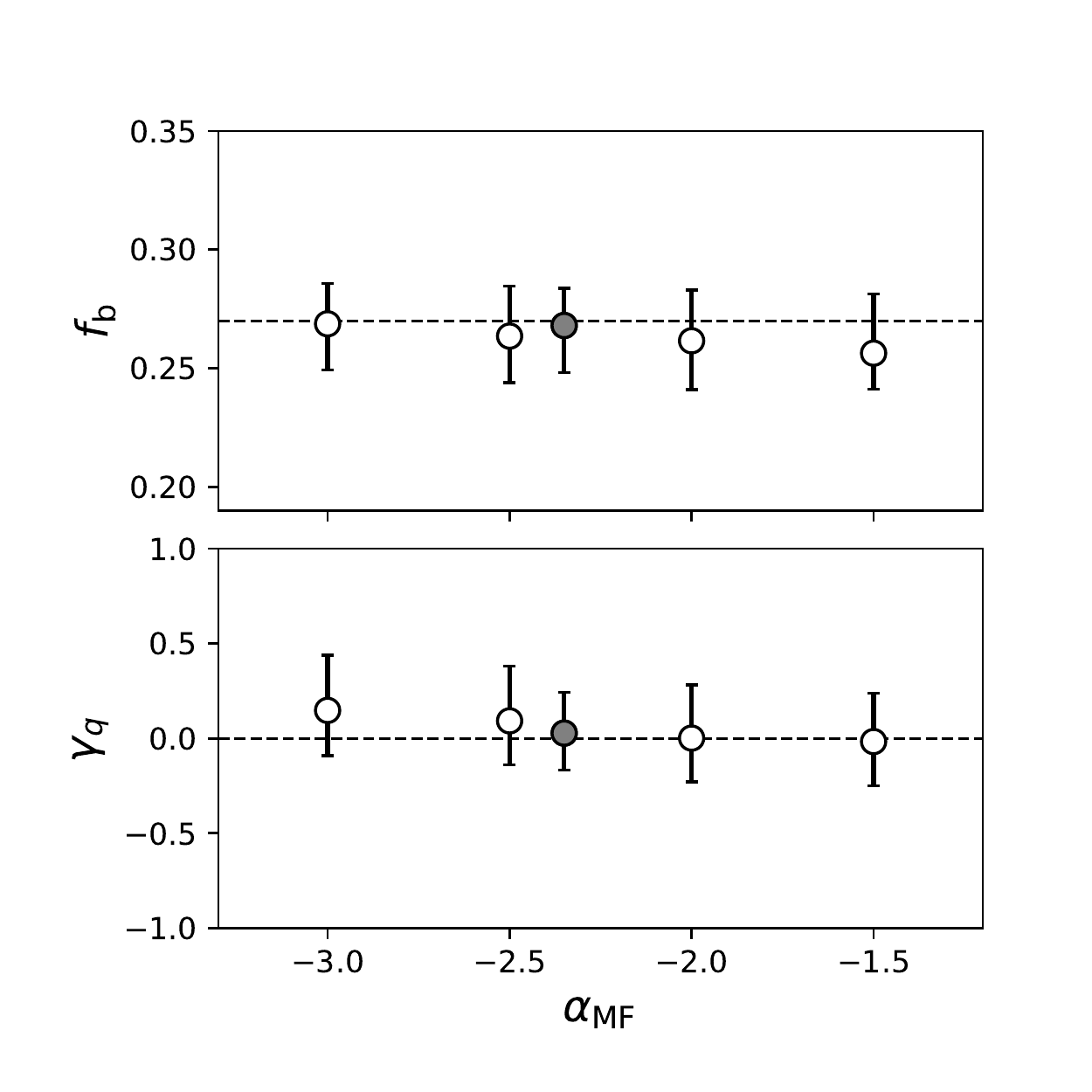}
  \caption{Differences between the inferred $\fb$ , $\gq$ and their true value as a function of $\alpha_\mathrm{MF}$.
  While in the fitting, the model mass function is forced to $\alpha_\mathrm{MF}=-2.35$. 
  }
  \label{fig:dMF-rst}
\end{figure}

As shown in Figure~\ref{fig:dMF-rst}, the difference in $\alpha_\mathrm{MF}$ will not affect the fitting results of $\fb$. For $\gq$, when the true underlying $\alpha_\mathrm{MF}$ is larger than the one used for model construction ($-2.35$), it will be underestimated. Anyway, this effect is not significant since the true value can still be covered by fitting uncertainties. Therefore, we conclude that the impact of mass function of the cluster is negligible in our fitting results.\\

\subsection{Reliability of MSRL}
\label{sec:Mock:MSRL}

As we have claimed in Section~\ref{sec:Mock:rst} and shown in Table~\ref{tab:mock-rst}, the small changes of location or extension of MS may largely bias the fitting result, while generally, the theoretical isochrone may not perfectly match the MS of a real cluster. So we have to check whether the MSRL determined by the Gaussian process (Section~\ref{sec:method:EMS:GP}) is good enough to substitute for the given MS of mock clusters. As a comparison, a statistics of 100 mock clusters shows that, along the whole MS, the average difference and its dispersion are all at the $10^{-4}$ mag level, which is significantly less than the observational error and the grid size of the digitalized model distribution.

Subsequently, we use the MSRL instead of the PARSEC isochrone in the fitting of the mock clusters, as listed in Table~\ref{tab:mock-rst} (case (h)). The resulting difference from the fiducial case is almost negligible comparing to the statistical uncertainty.

To sum up, the mixture model approach we established can well recover the true binary parameters, while the MSRL determined by the Gaussian process is good enough to take place of the real MS.
\\

\section{NGC3532 as an example}
\label{sec:3532}

In this section, we apply our method to infer the binary properties of the open cluster NGC3532 as a demonstration. In the following, we will first introduce the basic features of this cluster and the observational data. Then we derive the MSRL and its extension, and check for the mass function slope as well. Subsequently, the binary properties, together with its dependence on the stellar luminosity/mass and central radius, will be investigated through our mixture model approach. \\
 \begin{figure*}[!htbp]
  \centering
  \includegraphics[width=1\textwidth]{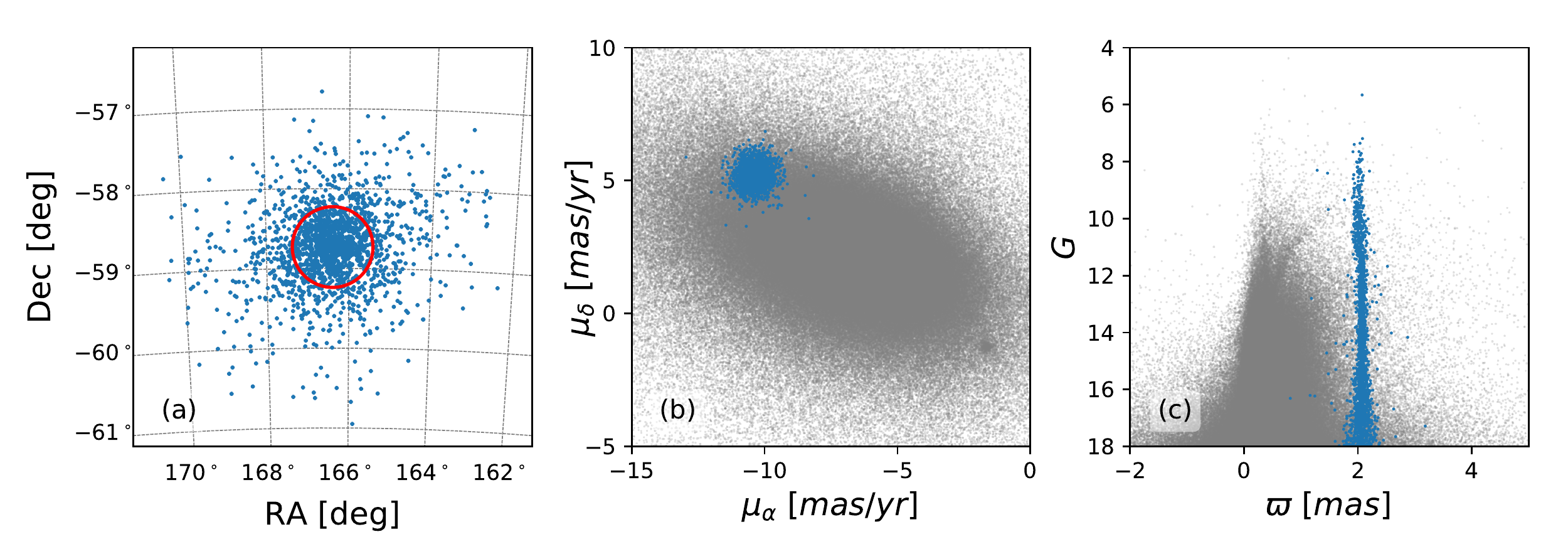}
  \caption{
  Distributions of stars in the region of NGC3532. Member stars are colored in blue, and the field stars are colored in gray. (a) Spatial distribution of cluster members (1879 stars) determined by \citet{GaiaCollaboration2018f}. The red circle represents the half number radius $r_\mathrm{h}$ of the {\it main sample} that consists of stars within 11.32 to 17.62 mag (see \ref{sec:3532:MainSamp}). (b) Vector point diagram illustrating the \gaia proper motions of stars within $2.3^\circ$ (the largest distance from the cluster center to the furthest member star) of NGC3532. (c) Same stars as panel (b) but showing the parallax versus $G$ magnitude.}
  \label{fig:basic-3532}
\end{figure*}

\subsection{General properties of NGC3532}
\label{sec:3532:gener}

NGC3532 is a middle-aged rich open cluster close to us in the southern sky ($\alpha_0=166.3975^\circ, \delta_0 = -58.7335^\circ$), and visible by the naked eye. Its distance is $\sim 485$ pc, and age is $\sim 400$ Myr, with the solar metallicity and quite small dust extinction (see Table~\ref{tab:3532-basic}). The mass function slope is very similar to the standard value of $\alpha_{\rm MF}$ for a wide mass range \citep{Clem2011b}. It has a clear MS and a prominent BS in the CMD, which makes it ideal for studying the binary population.

It is worth mentioning some other recent works that gave different measurements of its age and distance (\citet{Mowlavi2012}, \citet{Dobbie2009}, and \citet{Fritzewski2019}). These differences will not affect our measurement on binary properties since the fiducial MS we will use is not from their related theoretical isochrones, but the MSRL directly determined from the real data on the CMD.\\

\begin{table}[t]
\caption{Parameters of NGC3532.
}
\begin{center}
\begin{tabular}{lcl}
\hline\hline
 \quad Parameter      & Values &  Reference\\
\midrule{}
 \quad $\log (\mathrm{Age/yr})$      &   8.60 & \citet{GaiaCollaboration2018f} \\%
 \quad $ [\mathrm{Fe/H}]$       &   0.00 &\\%
 \quad DM(mag)     &   8.43  &\\%
 \quad $E(B-V)$            &   0.02 &\\%
\midrule
 \quad $\alpha_\mathrm{MF}$   &    $-2.39$ & \citet{Clem2011b}\\ %
 \quad                        &    $-2.35$ & Adopted by this work\\ %
\hline \hline
\end{tabular}
\label{tab:3532-basic}
\end{center}
\tablecomments{--- The age, metallicity, distance and extinction are taken from Table 2 of \citet{GaiaCollaboration2018f}. The mass function index is estimated by \citet{Clem2011b} and the value of ~\cite{Salpeter1955} is adopted in this study.}
\end{table}

\subsubsection{Cluster members and their {\gaia} photometric data}
\label{sec:3532:gener:data}

We use the membership catalog of \citet{GaiaCollaboration2018f}, which is mainly determined by accurate astrometric data of the {\gaia} DR2. It contains 1879 members with a flux limitation of $G=18$ mag. Within this, the $\BPRP$ color's observational errors are less than 0.01 mag for brighter members and consequently increase to $\sim 0.035$ mag at the faint end.

\reffig{fig:basic-3532} shows the distributions of coordinates, proper motions, parallax, and $G$ magnitudes for the cluster members. We can find very clear concentrations of member stars in the astrometric data spaces, extremely in the proper motion space, which makes the membership determination with very high confidence.\\

\subsubsection{MSRL and its extension}
\label{sec:3532:gener:MSRL}

\begin{figure*}[!htbp]
  \centering
  \includegraphics[width=1\textwidth]{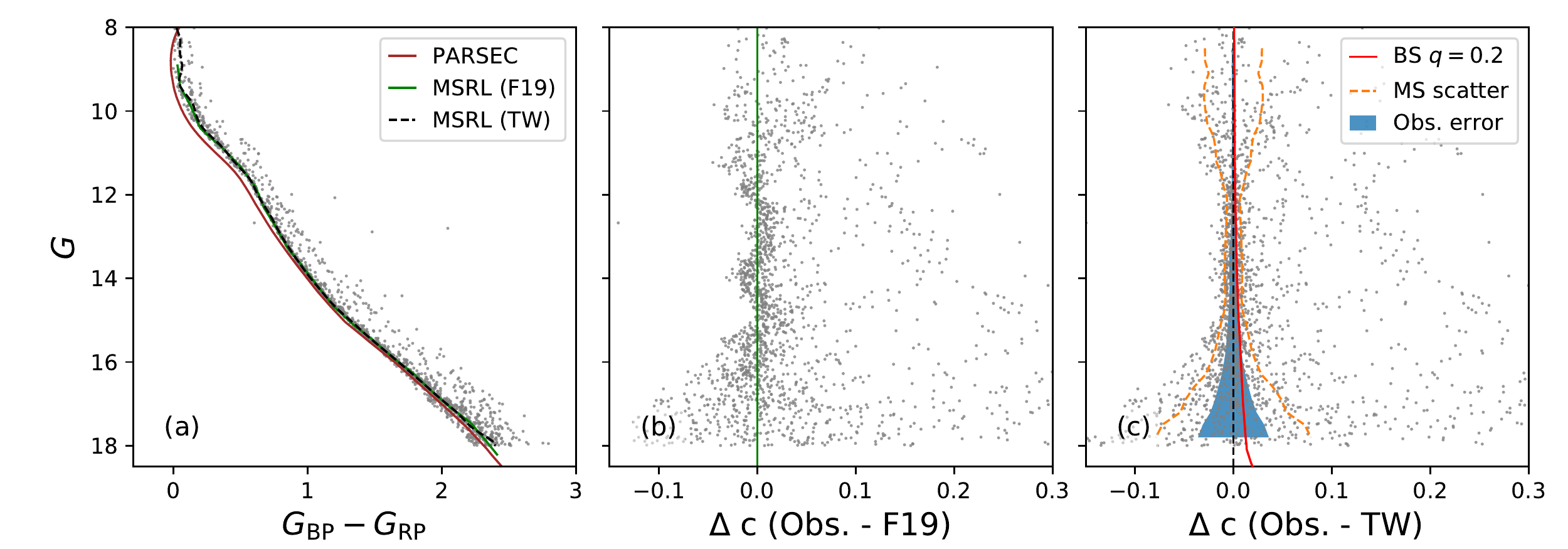}
  \caption{
  Comparison of the main sequence ridgelines (MSRL) and the scatters of NGC3532. (a) The color-magnitude diagram of the cluster. Member stars are shown as gray dots. The brown curve shows the PARSEC isochrone provided by G18, the green curve, and the black dash line show the MSRLs from \citet[F19]{Fritzewski2019} and the Gaussian process (this work, TW), respectively. (b) The residual color with MSRL by F19. (c) The residual color with MSRL by this work. Orange curves show the $1\sigma$ interval of residual color for single stars as a function of magnitude. The Blue shaded region shows the typical color uncertainties of \gaia measurement for comparison.
  The red curve shows the BS of $q=0.2$.
  }
  \label{fig:3532-MSRL}
\end{figure*}

As elaborated in Section~\ref{sec:method} and ~\ref{sec:Mock}, an accurate MS in the CMD is crucial for model construction. As illustrated in Figure~\ref{fig:3532-MSRL} panel(a), the theoretical PARSEC isochrone with cluster parameters provided by G18 (Table~\ref{tab:3532-basic}) shows a small but significant deviation from the visible MS. A similar discrepancy between observation and multiple isochrone models (including PARSEC, YaPSI, BaSTI, BHAC, MIST) has also been reported by \citet{Fritzewski2019}. The origin of the deviation remains unclear. It is probably caused by the inherent limitation of isochrone models for low-mass stars \citep{Khalaj2013a} or unknown defects in the {\gaia} color calibration, which is beyond the scope of this paper.

Such an amount of deviation might not significantly affect the inference of cluster properties, like the age, because it is mainly constrained by the overall shape of the isochrone. However, as discussed in Section~\ref{sec:Mock:rst}, the binary properties are susceptible to the precise location of the MS, imperfect isochrone model can severely bias the estimation of $\fb$ and $\gq$. An alternative solution is to employ the empirical MS instead.

Using the Gauss process introduced in Section~\ref{sec:method:EMS}, the MSRL of NGC3532 was derived as provided in Table~\ref{tab:MSRL} and plotted in the black dash line in \reffig{fig:3532-MSRL} as well. As shown in the figure, the ridgeline matches the observational data very well, even better than the RL manually made by \citet{Fritzewski2019}.

\begin{table}[htbp]
\caption{
  The empirical ridge line and scatters of the NGC3532 main sequence.
}

\begin{center}
\setlength{\tabcolsep}{6pt}
\begin{tabular}{cccccc}
\midrule \midrule
 \G & Mass & $ c_{\rm PAR}$ & $c_{\rm MSRL}$ & $\Delta c$ &$\sigma_c$ \\
 mag& $M_{\odot}$  & mag & mag &  mag & mag  \\

\midrule
10.000 & 2.015 & 0.066 & 0.161 & 0.095 & 0.028 \\
10.250 & 1.896 & 0.109 & 0.188 & 0.078 & 0.027 \\
10.500 & 1.785 & 0.166 & 0.245 & 0.079 & 0.022 \\
10.750 & 1.684 & 0.236 & 0.328 & 0.092 & 0.019 \\
11.000 & 1.593 & 0.315 & 0.399 & 0.084 & 0.018 \\
11.250 & 1.511 & 0.396 & 0.463 & 0.068 & 0.017 \\
11.500 & 1.435 & 0.467 & 0.537 & 0.070 & 0.013 \\
11.750 & 1.365 & 0.523 & 0.590 & 0.067 & 0.010 \\
12.000 & 1.300 & 0.571 & 0.628 & 0.056 & 0.008 \\
12.250 & 1.239 & 0.616 & 0.673 & 0.057 & 0.007 \\
12.500 & 1.181 & 0.664 & 0.721 & 0.057 & 0.007 \\
12.750 & 1.127 & 0.712 & 0.764 & 0.052 & 0.007 \\
13.000 & 1.076 & 0.764 & 0.807 & 0.043 & 0.008 \\
13.250 & 1.027 & 0.819 & 0.856 & 0.037 & 0.008 \\
13.500 & 0.981 & 0.876 & 0.908 & 0.033 & 0.009 \\
13.750 & 0.938 & 0.935 & 0.963 & 0.028 & 0.009 \\
14.000 & 0.896 & 0.996 & 1.022 & 0.026 & 0.009 \\
14.250 & 0.858 & 1.057 & 1.087 & 0.030 & 0.009 \\
14.500 & 0.822 & 1.123 & 1.153 & 0.030 & 0.008 \\
14.750 & 0.788 & 1.193 & 1.228 & 0.035 & 0.009 \\
15.000 & 0.756 & 1.266 & 1.315 & 0.049 & 0.012 \\
15.250 & 0.728 & 1.359 & 1.401 & 0.042 & 0.014 \\
15.500 & 0.702 & 1.453 & 1.489 & 0.035 & 0.017 \\
15.750 & 0.678 & 1.550 & 1.583 & 0.033 & 0.019 \\
16.000 & 0.653 & 1.647 & 1.676 & 0.029 & 0.023 \\
16.250 & 0.629 & 1.735 & 1.766 & 0.031 & 0.027 \\
16.500 & 0.605 & 1.822 & 1.856 & 0.034 & 0.036 \\
16.750 & 0.581 & 1.909 & 1.943 & 0.034 & 0.045 \\
17.000 & 0.557 & 1.995 & 2.035 & 0.040 & 0.050 \\
17.250 & 0.533 & 2.081 & 2.125 & 0.044 & 0.057 \\
17.500 & 0.508 & 2.164 & 2.200 & 0.035 & 0.072 \\
17.750 & 0.482 & 2.239 & 2.309 & 0.071 & 0.078 \\
18.000 & 0.457 & 2.311 & 2.396 & 0.084 & 0.079 \\
\midrule \midrule
\end{tabular}
\end{center}
\tablecomments{---The stellar mass and model color $c_{\rm PAR}$ as a function of $G$ magnitude are derived using the PARSEC isochrone with parameters specified in Table 2. The color of the main sequence ridge line $c_{\rm MSRL}$ and its scatter $\sigma_c$ are determined in this work. The differences between the isochrone and the ridge line $\Delta c = c_{\rm PAR} - c_{\rm MSRL}$ are also listed.
 }
\label{tab:MSRL}
\end{table}

The extension of the MSRL of NGC3532 is also determined by computing the scatters as described in Section~\ref{sec:method:EMS:ext}. These values are listed in Table~\ref{tab:MSRL} and also shown in Figure~\ref{fig:3532-MSRL}. Overall, the measured scatters are much larger than the observational errors reported by {\gaia} DR2.

The cluster members can be separated into two parts according to the relations between the MS scatters and the color errors. For stars fainter than $\sim 11.50$ mag, the scatter monotonously increases with both magnitude, and sharing the same trend with observational error of color. This phenomenon could be attributed to the observation issues, i.e., the underestimate of the $G_{\rm BP}$ and $G_{\rm RP}$ errors, or the additional noise between the calibration of these two passbands \footnote{Strictly speaking, $G$ band magnitude also suffers from such kind of scatter. However, the model is not sensitive to scatter in magnitude at all. We take $\sigma_G=0.01$ for simplicity.}. For the brighter stars, unfortunately, the scatters' change is opposite, with brighter stars having more extensive scatters. These scatters largely deviate from the small values of the observational errors. It implies that other intrinsic physics, e.g., the stellar rotation, may extend the MS near the turn-off point~\citep{Li2014}. These factors may complicate the model of MS and are beyond the scope of this paper. For this reason, we will omit the brighter members in the following analysis.

\subsubsection{The main sample}
\label{sec:3532:MainSamp}

Here we define the {\it main sample} of the NGC3532 members to investigate the binaries of this cluster. It has the flux limitations of $[11.32,17.62]$ mag, corresponding to $[0.5,1.5]\,M_\odot$ of single stars and $[0.42,1.29]\,M_\odot$ for equal-mass binaries. This selection avoids the problem of the abnormal extension of the MS in the brighter part. Moreover, we further rule out a few outliers with three times of $\sigma_c$ bluer than the MSRL or redder than the equal-mass BS. Finally, we get a {\it main sample} containing 1403 member stars. Please note that, for the {\it main sample}, the binaries we will discuss are around the solar mass. Actually, it covers the full FGK dwarfs' range.

Besides, we will further divide the {\it main sample} into two equal number disjoint subsets based on their magnitudes, or the radius from the cluster center, to investigate mass or radius dependences of the binary properties.\\

\subsubsection{Mass function}
\label{sec:3532:MF}

\begin{figure}[!htbp]
  \centering
  \includegraphics[width=0.45\textwidth]{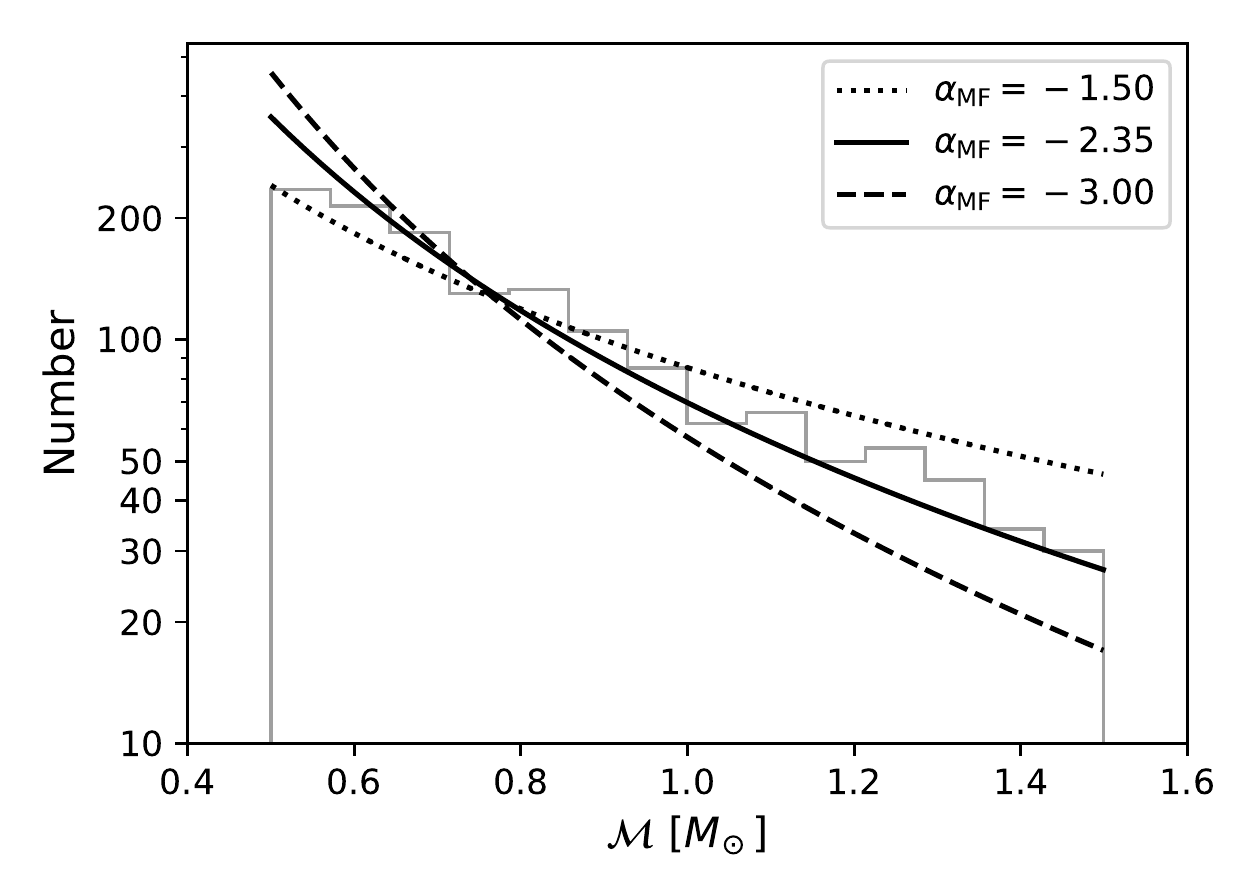}
  \caption{The stellar mass function of NGC3532. The histogram shows the mass function for single stars with $0.5M_\odot < \mathcal{M} <1.5 M_\odot$. Smoothed curves show the power-law distribution with various $\alpha_{\rm MF}$ for comparison.
  }
  \label{fig:3532-MF}
\end{figure}

We estimate the mass function of NGC3532 by counting the number of stars bluer than the MSRL since they are supposed to represent half of the single stars at any given mass corresponding to the mass to light relation from PARSEC. The result is shown as a histogram in Figure~\ref{fig:3532-MF}. Comparing with lines of different power index, the mass function of NGC3532 for single stars of the {\it main sample} should be very similar to the shape of $\alpha_{\rm MF}\simeq -2.35$, which agrees with the result of \citet{Clem2011b} for the mass range from 0.2 to 3.0 $M_\odot$.

According to the mock test of Section~\ref{sec:Mock:MF}, the mass function dependence of $\fb$ and $\gq$ could be neglected if it is not an abnormal $\alpha_{\rm MF}$ value. Therefore, the $\alpha_{\rm MF}$ could be assumed as $-2.35$ safely in the following analyses of the binaries of NGC3532. \\

\subsection{Inferred binary properties}
\label{sec:3532:rst}

We apply our method to infer the binary fraction $f_\mathrm{b}$ and mass-ratio index $\gamma_q$ of NGC3532. The MSRL listed in Table~\ref{tab:MSRL} is used as the fiducial MS, while the mass function is assumed as $\alpha_{\rm MF}=-2.35$. All the fitting results are summarized in Table~\ref{tab:3532-rst}. Based on the \texttt{emcee} sampling records, we also derived additional results of $\fb^{0.5}$ and $\fb^{0.7}$ for reference (see Section~\ref{sec:method:MM:LH} for details).
\\

\noindent\textbf{Results of the main sample}

\begin{figure*}[!htbp]
  \centering
  \includegraphics[width=0.45\textwidth]{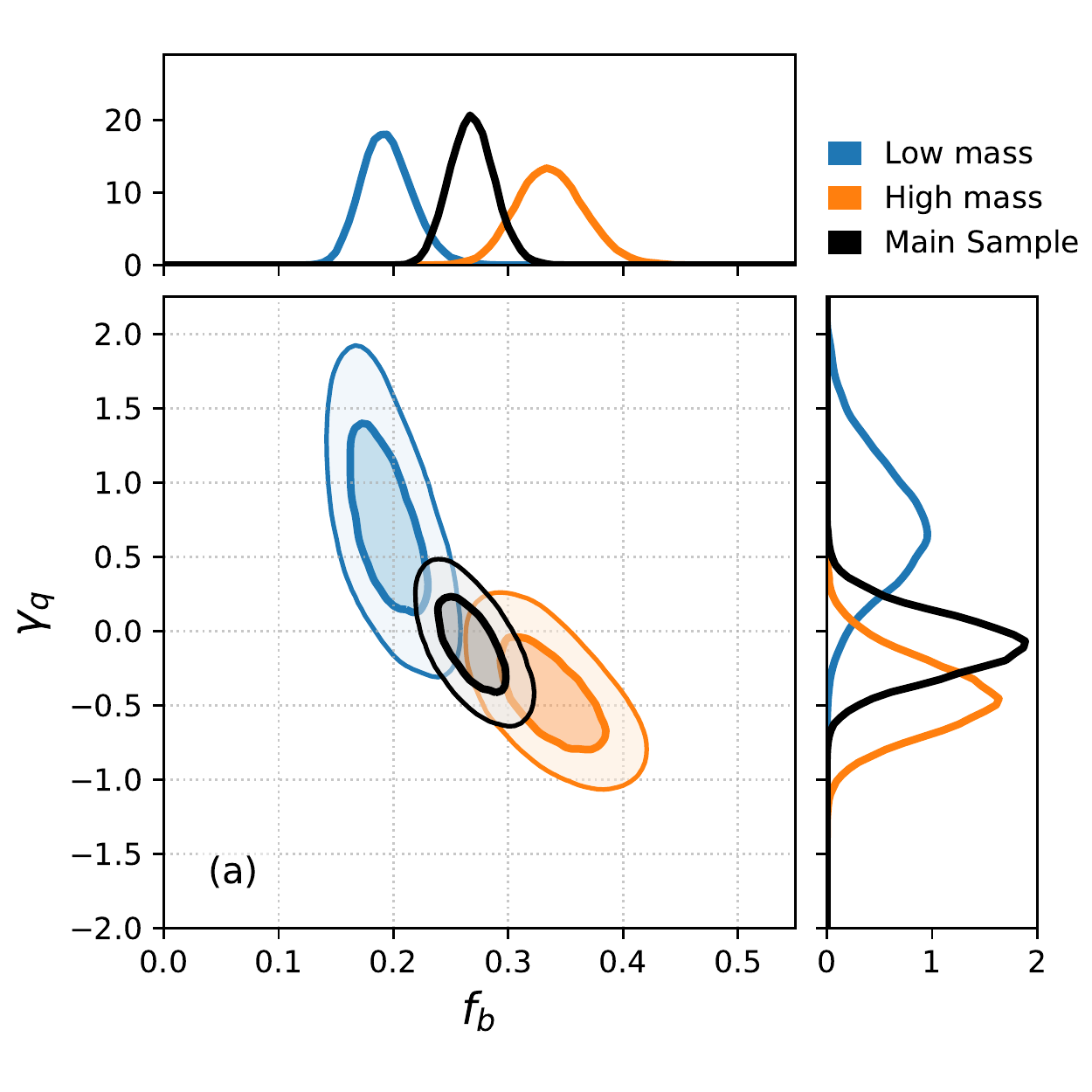}
  \includegraphics[width=0.45\textwidth]{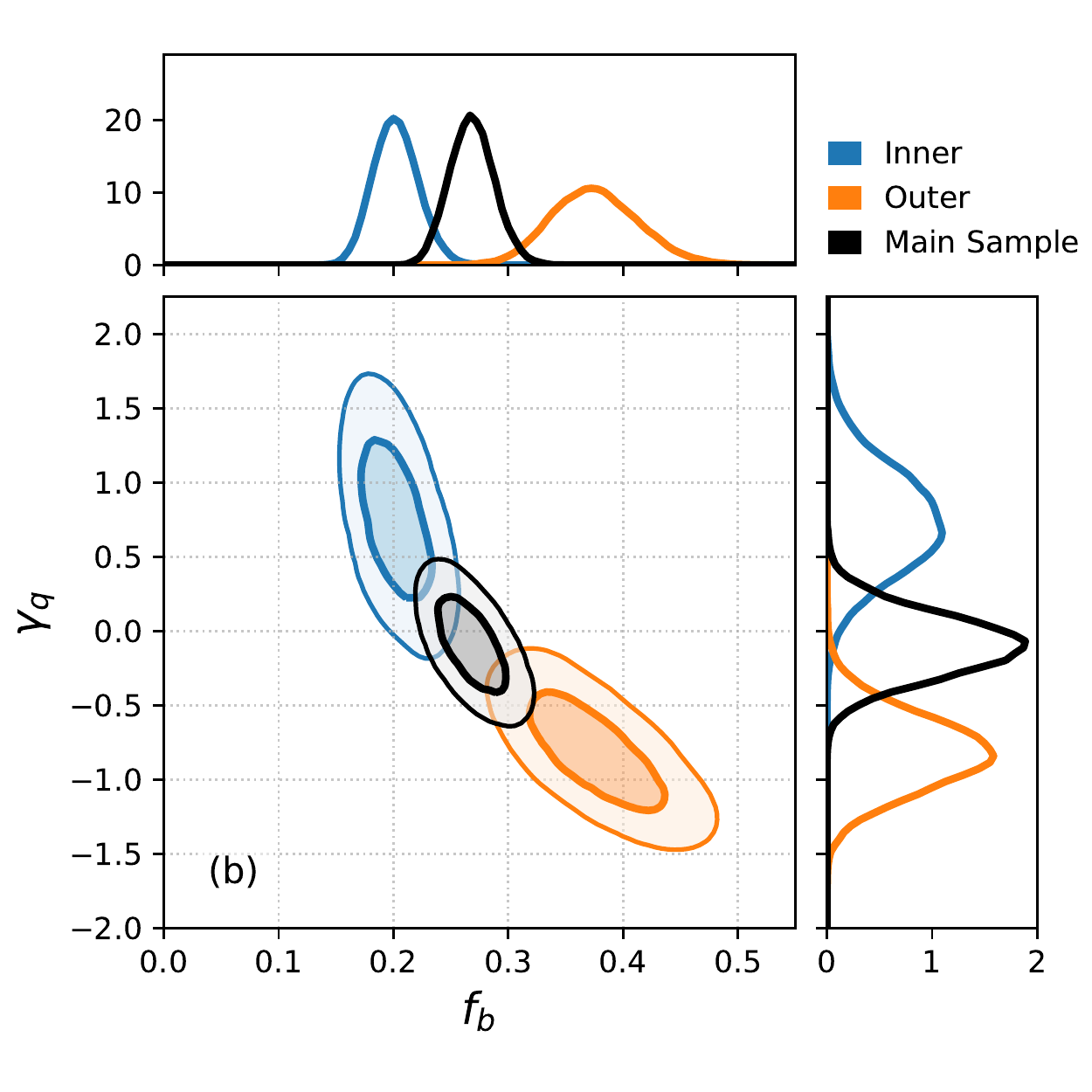}
  \caption{Probability density functions (PDFs) of parameters $\fb$ and $\gq$ based on the \texttt{emcee} sampling of NGC3532 for different sample stars. Each set of contours show the $1\sigma$ and $2\sigma$ confidence regions for $\fb$ and $\gq$ inferred from the relevant sample. The black contours are the same for two panels and show the distribution for the whole {\it sample}, whereas the blue (orange) contours show the distributions for the fainter (brighter) half of this sample in panel (a), and the inner (outer) part in panel (b). The corresponding marginalized PDFs for each sample are also shown as the colored curves in the top and right side panels.
  }
  \label{fig:3532-rst}
\end{figure*}

For the {\it main sample}, the two dimensional and marginal PDFs are shown in \reffig{fig:3532-rst} with black lines. A slight degeneracy between $\fb$ and $\gq$ are also found, similar to the fitting results of a mock cluster in \reffig{fig:mock-rst}. The uncertainty of $\fb$ is $\sim 0.02$, which is comparable to the dispersion from 100 mock clusters ($\sigma_{\fb}$ in Table~\ref{tab:mock-rst} ). It reaches a $\sim 7\%$ precision, which is a very high level in relevant works (e.g. $\sim20\%$ in \citet{Clem2011b}). The $\gq$ has a small negative value. Considering its fitting uncertainty, we can conclude that this cluster has an overall flat distribution of binary mass-ratio.\\

\noindent\textbf{Luminosity/mass dependence}
\label{sec:3532:massdp}

We further separate the {\it main sample} into two equal number subsets according to the brightness. The dividing magnitude is $G=15.06$, corresponding to $\mathcal{M}= 0.75 M_\odot$. This separation makes up two subsamples with different mass ranges. As we can find in the panel(a) of \reffig{fig:3532-rst} and Table~\ref{tab:3532-rst}, the difference between these two subsamples is obvious. There are more binaries for the massive subsample. Meanwhile, the more negative $\gq$ value indicates a non-flat $\mathcal{F}_q$ having more low mass-ratio binaries. The situation is just on the contract for the lower mass subsample. However, when we look at the other two derived parameters, $\fb^{0.5}$ and $\fb^{0.7}$, one may find that the fractions are getting converge. It seems that the difference of their $\gq$ is mainly caused by the different number of low mass-ratio binaries.\\

\noindent\textbf{Radius dependence}

When the {\it main sample} is separated by the half number radius $r_\mathrm{h}=30.35$ arcmin (4.28 pc), the two subsamples show a more significant difference in the panel (b) of Figure~\ref{fig:3532-rst} and Table~\ref{tab:3532-rst}. The inner region of NGC3532 has less fraction of binaries with a larger value of $\gq$, which means they lack the low mass-ratio binaries in this region. On the other hand, the outer region has much more low mass-ratio binaries, which leads to a higher value of $\fb$ and a more negative value of $\gq$. Similar to the case of the luminosity/mass dependence, the values of $\fb^{0.5}$ and $\fb^{0.7}$ are getting converge for these two regions.
\\

In summary, our method provides the highest precision in determining the binary fraction. Even in the cases of the subsamples with half number of member stars, the uncertainties can reach the $10\%$ level. Another interesting thing is that we can find the significant luminosity/mass dependence and radius dependence of binary properties, which may benefit from our method's ability to detect the extremely low mass-ratio binaries down to $q_{\rm min}=0.2$ . Otherwise, if we only consider the higher mass-ratio binaries, like $\fb^{0.5}$, these relations may too weak to be revealed.

It should be noted that there is a slight degeneracy between $\fb$ and $\gq$ (see the black contours in Figures~\ref{fig:single-mcmc} and~\ref{fig:3532-rst}), which is caused by the model and the statistical method we adopted. However, as we can see in Figure~\ref{fig:3532-rst}, the differences between subsamples are obviously, and definitely can not be attributed to the degeneracy. So we can claim that these mass and radius dependences are all substantial.\\

\begin{table*}[htbp]
\caption{
  Inferred binary fraction $\fb$ and mass-ratio index $\gq$ for NGC3532.
}
\begin{center}

\setlength{\tabcolsep}{15pt}
\begin{tabular}{cccccc}
\midrule \midrule
  & $N$ & $\fb$ & $\gq$ & $f^{0.5}_\mathrm{b}$ & $f^{0.7}_\mathrm{b}$\\
\midrule
 main sample       & 1403 & $0.267 \pm 0.019$ & $-0.10 \pm 0.22$ & $0.162 \pm 0.009$ & $0.096 \pm 0.007$\\
\cmidrule(lr){1-6}
    $  G < 15.06$ mag & 701 & $0.337 \pm 0.030$ & $-0.44 \pm 0.25$ & $0.182 \pm 0.014$ & $0.103 \pm 0.010$\\
    $  G > 15.06$ mag & 702 & $0.194 \pm 0.022$ & \phantom{$-$}$0.71 \pm 0.44$ & $0.143 \pm 0.012$ & $0.094 \pm 0.010$\\
\cmidrule(lr){1-6}
    $r < r_\mathrm{h}$ & 701 & $0.201 \pm 0.019$ & \phantom{$-$}$0.75 \pm 0.35$ & $0.149 \pm 0.013$ & $0.099 \pm 0.010$\\
    $r > r_\mathrm{h}$ & 702 & $0.374 \pm 0.037$ & $-0.82 \pm 0.26$ & $0.174 \pm 0.013$ & $0.092 \pm 0.010$\\
\midrule \midrule
\end{tabular}
\label{tab:3532-rst}
\end{center}
\tablecomments{$\fb^{0.5}$ and $\fb^{0.7}$ are derived variables rather than the original fitting parameters, represent the fraction of the binaries with $q>0.5$ and $q>0.7$. See Section~\ref{sec:method:MM:LH}.
 }
\end{table*}

\section{Discussion}
\label{sec:discu}

\subsection{The $f_b$ }
\label{sec:discu:fb}

\citet{Clem2011b} used the CMD of $B$ and $V$ photometric data to derived the $\fb$ of NGC3532 within a wide mass range from 0.2 to 3.0 $M_\odot$. They count the stars fainter than the MS and treat them as precisely half of the single stars. This method was simple but efficient, and avoided the problem of measuring the MS extension, though it is unable to model the binary mass-ratio distribution. They found $\fb =0.27\pm0.05$. This value is in perfect agreement with our result, but have less precision than ours ($\pm0.02$). It is a bit coincidence since the determined member stars and mass ranges of these two works are not exactly the same. However, the consistency of $\fb$ values partly verified that our method provides a {\it global} measurement, though we only use the {\it main sample} around the solar mass.

Considering the $q_{\rm min}=0.2$ we adopted, and the roughly flatten distribution of the mass-ratio ($\gq \sim 0$), one may infer that even if the very lower mass-ratio binaries ($q<0.2$) are taken into account, the $\fb^{tot}$ should be $\sim 0.32$. It is not a large value compared to the majority of OCs, especially for embedded clusters or star-forming regions, which often has more than 50 percent of binaries~\citep{Duchene2013}.

For the $\fb$ of field stars, it is found that there is a definite correlation between stellar mass and $\fb$ with higher mass subsets have more binary fractions, which is the same trend as of the NGC3532. But, for the similar mass range the FGK dwarfs, the $\fb$ of NGC3532 is significantly smaller than that of field stars \citep{Raghavan2010}.

Recently, \citet{Tian2018} estimated the binary fractions of 0.15 million dwarf stars observed with the LAMOST (DR4) and found $\fb$ is about 50\% for solar-type stars. \citet{Liu2019a} investigates the binary properties of field stars in the solar neighborhood using the LAMOST and {\gaia} data. His work focused on the stellar mass from 0.4 to 0.8 $M_\odot$, comparable to our lower mass subsample of NGC3532. He found a correlation between $\fb$ and $\gq$ for various subsamples divided by metallicity and stellar mass. The fitting results of his low $\gq$ subsets ($\gq \sim 0$) show that the lower limit of $\fb$ is about $0.2$, which is similar to our lower mass subsample, $\fb = 0.194\pm0.022$.

In short, the binary fraction of NGC3532 is on the low side of normal.
\\

\subsection{The $\gq$}
\label{sec:discu:gamma}

The mass-ratio distribution $\mathcal{F}_q$ is a critical diagnose of the binary formation models, whether it is a tidal capture or a fragmentation process. The tidal capture model predicts that for each primary star, the mass of its secondary star is chosen randomly from the single-star mass function, so the $\mathcal{F}_q$ reflects the $\mathcal{F}_{\rm MF}$ with $\gq \sim \alpha_{\rm MF}$. On the other hand, the fragmentation scenario contains multiple steps. It assumes that a binary system firstly forms by fragmentation of either the same collapsing molecular cloud core or a newly formed star-disk. Subsequently, other dynamical processes, such as the disk accretion and star interaction, will contribute to determining the properties of binary. Therefore the final profile of the mass-ratio distribution of binaries may significantly diverge to the shape of the single-star mass function. Generally, we found the mass-ratio distribution of NGC3532 is flattened ($\gq \sim 0$). Even for the subsets with lower $\gq$ values, i.e., $\gq = -0.438$ for the higher mass and $\gq = -0.818$ for the outer region, they are all significantly more flatten than the slope of the mass function ($\alpha_{\rm MF} \sim -2.35$). These results rule out the simple tidal capture model.

We have to note that the uncertainty of $\gq$ is relatively large. It is also worth to point out that $\gq$ is not a robust variable, which is sensitive to the fitting method~\citep{Reggiani2013}. Taking these factors in to account, our results of $\gq$ are generally in good agreement with previous works, e.g., $\gq \sim 0.0$ for binaries in OB association \citep{Kouwenhoven2007a}, and for field stars, $\gq = 0.25\pm 0.29$ for solar type binaries ~\citep{Reggiani2013}, $0.3\pm0.1$ for FGK dwarfs \citep{Duchene2013}, and $-1.0$ -- $\sim$3.0 for various subsets based on metallicity and stellar mass \citep{Liu2019a}.

\citet{Fisher2005} and \citet{El-Badry2019} argue that the mass-ratio of binary stars does not conform to the single power-law distribution. It has an overall flatten distribution or a broken power law but peaked at $q=1$. Moreover, \citet{Kouwenhoven2007a} also discussed a Gaussian distribution. Nevertheless, even if we force a power-law shape to fit the parameters, the result of $\gq$ can still reflect this situation with an undoubted positive fitting value. For NGC3532, this situation may show in the cases of lower mass or inner region subsets, which can be attributed to the dynamical process, that will be discussed in the next subsection. \\

\subsection{Mass or radius dependences and their dynamical implication}
\label{sec:discu:dynam}

In Section~\ref{sec:3532:rst}, we have verified that there are stellar mass and radius dependences of binary properties of NGC3532. The lower mass subsample has fewer number of binaries but a larger value of $\gq$. Meanwhile, the inner cluster region also tends to have fewer binaries and the steeper mass ratio distribution towards the equal-mass binaries. Interestingly, the differences between these subsets can be attributed to only one factor, the lack of low mass-ratio binaries for specific subsets. That means, if there are mechanisms to reduce the fraction of low $q$ binaries of the lower mass subsample or the inner region subsample, then the mass and radius dependences will be expressed.

It is reported that both $\fb$ and $\gq$ are correlated to the metallicity (e.g., \citealt{Liu2019a}). However, as an OC, the members of NGC3532 are supposed to have similar metallicities. Moreover, considering that the age of NGC3532 is 400 Myr, which is much longer than its relaxation time $\sim100$ Myr, this cluster has undergone a roughly enough dynamical evolution. It seems reasonable to suspect that the current properties of binaries are very different from their primordial state. Therefore, the main reason for the mass and radius dependences is the internal dynamics of the cluster.

The interaction in a cluster can lead to two possible endings for a binary. The most violent one is to disrupt the binary systems directly. Since the binary binding energy $E_b \propto q\mathcal{M}^2$, a system with a smaller primary star or a lower mass-ratio will get disrupted more efficiently. In other words, the higher primary mass or the equality mass-ratio helps the survival of a binary during the encounter. It could explain the current situation where the high mass subsample keeps a larger fraction of binaries, while the low mass subsample has lost lots of low $q$ binaries. This trend is consist with \citet{Kaczmarek2011} and \citet{Dorval2017} in young dense clusters. Nevertheless, it might not be universal since, for example, an opposite trend was reported in $\alpha$ Persei and Praesepe\citep{Patience2002a}.

Meanwhile, it could be intuitive to understand that in the inner region, where the interactions happen frequently, more low $q$ binaries are disrupted than those in the outer region. This phenomenon is consistent with that of globular clusters \citep{deGrijs2013}, as the binary fraction is smaller in the dense region.

\citet{Parker2013} claimed that the shape of the mass-ratio distribution is an outcome of the star formation process, rather than a dynamical result. Unfortunately, their simulation only lasts 10 Myr, which is too short for the dynamical process to change the mass-ratio distribution. We have to argue that the different $\gq$s between various subsamples of NGC3532 is a piece of strong evidence for the influence of dynamics on the shape of the mass-ratio distribution. Because the interaction effect is related not only to the primary mass but also on the mass-ratio, after hundreds of Myrs accumulation of the interaction effect, the mass-ratio distribution shape would be distinguishable at different local environments.

Another possible ending for a binary experienced a dynamic encounter is that the binary system remains, but one of the companion stars (usually the secondary star) has been replaced by a more massive perturbing star. This process will enlarge the $q$ values of this system. If this process frequently happened and is more efficiently in the dense region, one may expect that there would be more equal-mass binaries in the inner region of a cluster than in its outer part. According to the approximately equal values of the $\fb^{0.7}$s of radius separation subsamples of NGC3532 (Table~\ref{tab:3532-rst}), it seems no significant evidence for this mechanism in this middle-aged cluster. Thus, one may conclude that the disruption might be the dominant dynamical effect in the early stage of a cluster when the replacement is still hard to be detected. This assumption could be further testified using an OC sample that covers a wide range of cluster age.

\section{Conclusion}
\label{sec:sum}


We have developed a comprehensive approach to model the mixture distribution of single stars and binaries of an open cluster in the color-magnitude diagram (CMD), which enables us to infer the binary properties accurately and precisely, in particular, the binary fraction $\fb$ and binary mass-ratio index $\gq$. We have further tested the validity and accuracy of the method with mock clusters.

By using the \gaia DR2 photometric data and the astrometric members, the open cluster NGC3532 is found to have $\fb=0.267\pm0.019$ and $\gq=-0.10\pm0.22$. These results imply that NGC3532 is not a binary-rich cluster, and its binary mass ratio follows a nearly uniform distribution.

We further unveil the mass and radius dependences of binary properties. The lower mass and the inner region stars have few low mass ratio binaries. Such correlations are evidence of internal dynamical interaction, consistent with the argument that binaries with smaller primary mass or lower mass-ratio get disrupted by interactions more efficiently due to their lower binding energy.

However, it should be reiterated that the discovery of such mass/radius dependences in OCs is based on the rigorous determination of the MS position and extension. Otherwise, the imprecise MS can severely bias the fitting parameters.

One may wonder whether this mass/radius dependences is a general phenomenon in OCs, or how important role is the dynamical interaction plays. It requires further study of large open cluster samples that cover broader ranges of mass, age, metallicity, and environment.

\acknowledgments
We sincerely thank the anonymous referee for valuable comments and suggestions. We thank Chao Liu, Xiaoying Pang for helpful discussions. This work is supported by National Key R\&D Program of China No.\ 2019YFA0405501. This work has made use of data from the European Space Agency (ESA) mission {{\gaia}} (\url{https://www.cosmos.esa.int/gaia}), processed by the {{\gaia}} Data Processing and Analysis Consortium (DPAC, \url{https://www.cosmos.esa.int/web/gaia/dpac/consortium}). Funding for the DPAC has been provided by national institutions, in particular the institutions participating in the {{\gaia}} Multilateral Agreement.


\software{Astropy \citep{AstropyCollaboration2013}, PARSEC \citep{Bressan2012}, scikit-learn \citep{Pedregosa2012}, Numpy \citep{vanderWalt2011}, Scipy \citep{Oliphant2007}, Matplotlib \citep{Hunter2007}, \texttt{emcee} \citep{Foreman-Mackey2013}}.
\bibliographystyle{aasjournal}
\bibliography{main}

\end{CJK*}
\end{document}